\documentclass
[11pt, final, report, onecolumn]{IEEEtran}
\pdfoutput=1 
\IEEEoverridecommandlockouts
\usepackage[cmex10]{amsmath}
\usepackage{amsfonts}
\usepackage{amssymb}
\usepackage{amscd}
\usepackage[dvips]{graphicx}
\usepackage{epsfig}
\usepackage{mathrsfs}
\usepackage{afterpage}
\usepackage{graphicx}
\usepackage{cite}
\usepackage{enumerate}
\usepackage{url}
\usepackage{xcolor}

\newcommand{\nn}{\nonumber\\}

\newcommand{\U}{{U}}

\newcommand{\E}{\mathbb{E}}
\newcommand{\N}{{N}}
\newcommand{\F}{\mathcal{F}}
\newcommand{\Fk}{\mathfrak{F}}
\newcommand{\G}{\mathcal{G}}

\newcommand{\ii}{\mathfrak{i}}
\newcommand{\M}{\mathcal{M}}
\newcommand{\as}{\text{a.s.}}
\newcommand{\tin}{t\in [0,T]}
\newtheorem{Theorem}{Theorem}
\newtheorem{Lemma}{Lemma}
\newtheorem{Def}{Definition}

\newcommand{\newadd}{}
\newcommand{\aaronadd}{}

\begin{document}
\title{The Stochastic-Calculus Approach to \\
Multi-Receiver Poisson Channels}
\author{\IEEEauthorblockN{Nirmal V. Shende} and
\IEEEauthorblockN{Aaron B. Wagner}
\thanks{The authors are with the School of Electrical and Computer Engineering,
Cornell University,
Ithaca, NY 14853. Email:\{nvs25,wagner\}@cornell.edu. This research was supported by the US National Science
Foundation under grants CCF-1065352 and CCF-1513858. This paper was presented  at the IEEE Int. Symposium on Information Theory (ISIT), Barcelona, July 2016.}
}
\maketitle
\begin{abstract}
 We study two-receiver Poisson channels using tools derived from stochastic calculus. We obtain a general formula for the mutual information over the Poisson channel that allows for conditioning and the use of auxiliary random variables. 
We then use this formula to compute necessary and sufficient conditions under which one Poisson channel is less noisy and/or more capable than another, which turn out to be distinct from the conditions under which this ordering holds
for the discretized versions of the channels.
We also use general formula to determine the capacity region of the more capable Poisson broadcast channel with independent message sets, the more capable Poisson wiretap channel, and the general two-decoder Poisson broadcast channel with degraded message sets.
\end{abstract}
\section{Introduction}

The Poisson channel models a direct-detection optical communication system in which the input to the channel $X_0^T$ represents the strength of the optical input signal, and the output of the channel is a Poisson process with rate $aX_0^T+\lambda$, where $a$ accounts for attenuation and $\lambda$ represents the rate of the dark current.
Capacity studies of this channel have been ongoing since it was introduced
as a viable model in~\cite{Personick,Mazo}.

Broadly speaking, the channel has been studied using two 
mathematical approaches. Early work calculated mutual information
and related quantities for the channel using stochastic calculus
and, in particular, the theory of point process martingales~\cite{Kabanov,Davis}.
Most later work followed the approach of Wyner~\cite{Wyner} who 
argued that the encoder and decoder could be restricted to use
the channel so that it behaves like a discrete-time, memoryless,
binary channel, with no essential loss of performance. One then
applies standard techniques for such channels~\cite{Shannon, Bergmans, Korner,Gamal1}.

We espouse the former approach in this paper, both on the general
principle that, when the existing tools are insufficient for a 
new problem, it is preferable to extend the tools rather than 
to reduce the problem, and for certain pragmatic reasons.
The reduction to a discrete-time binary channel is somewhat involved,
and it must be reproved for each new variation.
Once the appropriate 
stochastic-calculus-based tools have been developed, on the
other hand, they can be directly applied to new problems.
Moreover, it is unclear how to extend Wyner's~\cite{Wyner} 
reduction to some setups, such as the wiretap version of 
the channel considered herein.

Of course, the stochastic calculus approach also has its
disadvantages: it requires more sophisticated mathematics,
and one cannot apply results from the extensive literature
on discrete memoryless channels. One cannot even presume
that the capacity is governed by the maximal mutual 
information, for instance, an oversight in the early
work that used this approach. On the 
other hand, once the necessary tools are developed,
coding theorems follow expeditiously.

The goal of this paper is to develop those tools that
are necessary for various multi-decoder extensions
of the Poisson channel. The two-decoder Poisson channel consists of a single transmitter (which inputs process $X_0^T$) and two receivers with output processes $Y_0^T$ and $Z_0^T$, where $Y_0^T$ and $Z_0^T$ are Poisson process with rates $a_yX_0^T+\lambda_y$ and $a_zX_0^T+\lambda_z$, respectively.
 We shall consider
both the broadcast channel (either with independent or degraded message sets) and the wiretap channel (where one of the receivers is an eavesdropper). 

We derive a general formula for the mutual information
over a Poisson channel, which generalizes an existing formula~\cite{Kabanov, Davis}
by allowing the use of auxiliary random variables and conditioning.
We also obtain a continuous-time Csisz\'{a}r-sum-like identity for Poisson channels. Using these tools, we obtain necessary and sufficient conditions for which the broadcast channel is less noisy and more capable, and show that these orderings are in fact equivalent. These conditions turn out not to be equivalent, however,
to the analogous conditions for the discrete-time binary channel
obtained as a reduction of the Poisson channel~\cite{Kim}, indicating
that some care is required when interpreting results obtained via this
reduction.
We also rederive the capacity of the more capable broadcast channel with independent message sets (found earlier using the reduction method~\cite{Kim}), extend the secrecy capacity results of the degraded wiretap channel to the more capable wiretap channel, and obtain the capacity of the broadcast channel with degraded message sets.
\section{Preliminaries}
We will construct  a probability space $(\Omega,\F,P)$   on which all stochastic processes considered here are defined. For a finite $T>0$, let $(\mathcal{F}_t:\tin)$ be an increasing family of $\sigma$-fields with $\F_T\in\F$.     Stochastic processes are denoted as $X_0^T=\{X_t,0 \le t \le T\}$. \newadd{$X_{t-}$ denotes   $\lim_{\delta\to 0^{+}} X_{t-\delta}$ when $t>0$, and equals $X_0$ when $t=0$}.
The process $X_0^T$ is said to be \emph{adapted} to the  history $(\mathcal{F}_t:t\in [0,T])$ if $X_t$ is $\F_t$ measurable for all $t\in[0,T]$.
 The internal history recorded by the process $X_0^T$ is denoted by $\mathcal{F}^X_t=(\sigma(X_s): s \in [0,t])$, where $\sigma(A)$ denotes the  $\sigma$-field generated by  $A$.  
A process $X_0^T$ is called $(\F_t:t\in [0,T])$-\emph{predictable} if $X_0$ is $\F_0$ measurable and the mapping $(t,\omega)\to X_t(\omega)$ defined from $(0,T)\times\Omega$ into $\mathbf{R}$ (the set of real numbers) is measurable with respect to the $\sigma$-field over $(0,T)\times\Omega$ generated by rectangles of the form 
\begin{align}
(s,t]\times A; \quad 0< s \leq t \leq T, \quad A \in \F_s.
\label{EQ:Pred_field}
\end{align}
Let $\mathcal{N}_0^T$ denote the set of counting realizations (or point-process realizations) on $[0,T]$, i.e., if ${N}^T_0\in\mathcal{N}_0^T$, then for $t\in[0, T]$, ${N}_t\in \mathbf{N}$ (the set of non-negative integers), is right continuous, and has unit jumps with ${N}_0=0$. 

For two given   $\sigma$-fields $\mathcal{F}_1$ and $\mathcal{F}_2$, the smallest  $\sigma$-field containing the union of these two fields is denoted by $\mathcal{F}_1 \vee \newadd{\mathcal{F}_2 }$.  For two measurable spaces $(\Omega_1,\F_1)$ and $(\Omega_2,\F_2)$, the product space is denoted by $(\Omega_1 \times \Omega_2,\F_1\otimes\F_2)$. We say that $A \rightleftarrows B \rightleftarrows C$ forms a Markov chain under measure $P$, if $A$ and $C$ are conditionally independent given $B$ under $P$. $P\ll Q$ denotes that the probability measure $P$ is absolutely continuous with respect to the measure $Q$. $\textbf{1}\{\mathsf{E}\}$ denotes the indicator function for an event $\mathsf{E}$ and $\log(x)$ is the natural logarithm of $x$.   Convergence in probability and almost sure (a.s.) convergence are denoted by $\xrightarrow{\text{p}}$ and $\xrightarrow{\text{\as}}$, respectively. Throughout this paper we will adopt the convention that $0\log(0)=0$, $\exp(\log(0))=0$, and $0^0=1$. 

We will  use the following form of Jensen's inequality.
\begin{Lemma}
If $\phi(x)$ is a convex function, then 
\begin{align*}
\E[\phi(X)]\ge\E[\phi(\E[X|A,B])]\ge\E[\phi(\E[X|A])]\ge\phi(\E[X]).
\end{align*}
\end{Lemma}

We now recall the definition of mutual information for general ensembles and
its properties.
Let $A$, $B$, and $C$ be measurable mappings defined on a given probability space $(\Omega,\F,P)$, taking values in $(\mathcal{A},\mathfrak{F}^A)$, $(\mathcal{B},\mathfrak{F}^B)$, and $(\mathcal{C},\mathfrak{F}^C)$ respectively.  
Consider partitions of $\Omega$, $\mathfrak{Q}_A=\left\{\mathtt{A}_i, 1\le i \le N_A\right\}\subseteq\sigma(A)$ and $\mathfrak{Q}_B=\left\{\mathtt{B}_j, 1\le j \le N_B\right\}\subseteq\sigma(B)$. Wyner defined the conditional mutual information $I(A;B|C)$  as~\cite{WYNER197851} 
\begin{align}
I(A;B|C)=\sup_{\mathfrak{Q}_A,\mathfrak{Q}_B} \E\left[\sum_{i,j=1,1}^{N_A,N_B}P(\mathtt{A}_i,\mathtt{B}_j|C)\log\left(\frac{P(\mathtt{A}_i,\mathtt{B}_j|C)}{P(\mathtt{A}_i|C)P(\mathtt{B}_j|C)}\right)\right],
\label{EQ:MI_Def}
\end{align}
where the supremum is over all such partitions of $\Omega$. 
Wyner  showed that $I(A;B|C)\ge 0$  with equality if and only if $A \rightleftarrows C \rightleftarrows
 B $ forms a Markov chain \cite[Lemma 3.1]{WYNER197851}, 
and that  (generally referred to as) Kolmogrov's formula holds \cite[Lemma 3.2]{WYNER197851}
\begin{align}
I(A,C;B)=I(A;B)+I(C;B|A).
\label{EQ:Kolmogrov}
\end{align}
Hence if $I(A;B)<\infty$, then $I(C;B|A)=I(A,C;B)-I(A;B)$. The data processing inequality can be obtained from (\ref{EQ:Kolmogrov}) as well: if $A \rightleftarrows C \rightleftarrows
 B $ forms a Markov chain, then $I(A;B)\le I(C;B)$.

Denote by $P^{A,B}$, the joint distribution of $A$ and $B$ on the space ($\mathcal{A}\times \mathcal{B},\Fk^A\otimes\Fk^B$ ),  i.e.,
\begin{align*}
P^{A,B}(dA\times dB)=P((A^{-1}(dA),B^{-1}(dB)), \quad dA\in \Fk^A, dB\in\Fk^B.
\end{align*} 
Similarly, $P^A$ and $P^B$ denote the marginal distributions.
 Gelfand and Yaglom \cite{Gelfand} proved that   if $P^{A,B}\ll P^A\times P^B$, then the mutual information $I(A;B)$ (defined via (\ref{EQ:MI_Def}) by taking $\sigma(C)$ to be the trivial $\sigma$-field) can be computed as 
\begin{align}
I(A;B)=\E\left[\log\left(\frac{dP^{A,B}}{d(P^A\times P^B)}\right)\right].
\label{EQ:MI_LLR}
\end{align}
A sufficient condition for $P^{A,B}\ll P^A\times P^B$ is that $I(A;B)<\infty$ \cite[Lemma 5.2.3, p. 92]{Gray}.
We will also require the following result \cite[Lemma 2.1]{WYNER197851}:
\begin{Lemma}[Wyner's Lemma]
\label{Le:Wyner}
If $M$ is a finite alphabet random variable, then
\begin{align*}
I(M;U_0^T)=H(M)-\E\left[H(M|U_0^T)\right],
\end{align*}
where 
\begin{align*}
H(M|U_0^T)=-\sum_{m}P(M=m|U_0^T)\log\left(P(M=m|U_0^T)\right),
\end{align*}
\end{Lemma}
 and $H(M)$ is the entropy of $M$.

\section{Doubly-Stochastic Poisson Process}
\begin{Def}
\label{Def:DSPP}
Let $X_0^T$ be a non-negative process. A counting process $\N_0^T$  is called \emph{a doubly-stochastic Poisson process with rate process $X_0^T$}  under measure $P$ if 
\begin{itemize}
\item for an interval $[s
, t]\in[0,T]$
\begin{align*}
P(\N_t-\N_s=k|X_0^T)=\frac{1}{k!}\left(\int_s^t X_\tau \,d\tau\right)^k\exp\left(-\int_s^t X_\tau \,d\tau\right), \mbox{ for } k\in\mathbf{N} 
\end{align*}
with convention $0^0=1$,
\item conditioned on $X_0^T$ the increments in disjoint intervals of $[0,T]$ are independent.
\end{itemize}
\end{Def}
Throughout this paper, the rate process $X_0^T$ will be a bounded c\`adl\`ag  (right continuous with left limits) process.
\begin{Def}
If $N_0^T$ is a counting process adapted to the history $(\F_t:\tin)$, then $N_0^T$ is said to have  $(P,\F_t:\tin)$-\emph{intensity} ${\Gamma}_0^T=\{\Gamma_t, 0\le t \le T \}$, where $\Gamma_0^T$ is a non-negative measurable process if
\begin{itemize}
\item $\Gamma_0^T$ is $(\F_t:t\in [0,T])$-predictable,
\item $\int_0^T\Gamma_t\,dt < \infty$, $P$-a.s.,
\item and for all non-negative $(\F_t:\tin)$-predictable processes $C_0^T$:\footnote{The limits of the Lebesgue-Stieltjes integral $\int_a^b$ are to be interpreted as $\int_{(a, b]}$.}
\begin{align*}
\E\left[\int_0^T C_s \, d\N_{s}\right]=\E\left[\int_0^T C_s \Gamma_s \, ds\right]. 
\end{align*}
\end{itemize}
\end{Def}
\begin{Def}
Given a doubly-stochastic Poisson process $\N_0^T$, a counting process $\tilde{\N}_0^T$ is called  \emph{the time-reversed $\N_0^T$ process} if $\tilde{N}_0=0$ and for $t\in(0,T]$, $\tilde{N}_t=N_{T}-\N_{(T-t)-}$.
\end{Def}
\begin{Def}
Fix $0\le t_1 < t_2 \le T$. Given a doubly-stochastic Poisson process $\N_0^T$,  $\N_{t_1}^{t_2}$ will denote a point process on $[0, T]$ which has no arrival before $t_1$,  after $t_2$, and the same arrivals as process $\N_0^T$ on the interval $[t_1, t_2]$. Specifically, let $\hat{\N}_t$ denote the value of the process $\N_{t_1}^{t_2}$ at time $t$. Then 
\begin{align*}
\hat{\N}_t&=0, \quad t < t_1, \nn
&=N_t-N_{t_1}, \quad t_1 \le t\le t_2, \nn
&=N_{t_2}-N_{t_1}, \quad t_2 < t \le T. 
\end{align*}
\end{Def}
\begin{Lemma}
\label{LE:DSPP}
Suppose $\N_0^T$ is a doubly-stochastic Poisson process with rate process ${X}_0^T$ under measure $P$ and ${\tilde{\N}}_0^T$ is the time-reversed $\N_0^T$ process. Then ${\tilde{N}}_0^T$ is a doubly-stochastic Poisson process with rate process ${\tilde{X}}_0^T=\left\{\tilde{X}_t=X_{(T-t)-}: t\in [0, T]\right\}$ under measure $P$.
\end{Lemma}
\begin{IEEEproof}
See the Appendix. 
\end{IEEEproof}

\begin{Lemma}
\label{Le:MC_Int}
Suppose $N_0^T$ is a doubly-stochastic Poisson process with  rate process $\Lambda_0^T$ under measure $P$ and $A\rightleftarrows \Lambda_0^T\rightleftarrows N_{t_1}^{t_2}$ is a Markov chain. Let $\hat{N}_0^T=\{\hat{N}_t:\tin\}$, where $\hat{N}_t$ is the value of $N_{t_1}^{t_2}$ at time $\tin$, i.e., the process $\hat{N}_0^T$ has no arrivals prior to $t_1$ and after $t_2$ and the same arrivals instants as process $N_0^T$ for $t\in[t_1, t_2]$.  
Then for $\F_t=\sigma(A)\vee\F^\Lambda_T\vee\F_t^{\hat{N}}$, the $(P,\F_t:\tin)$-intensity of $N_0^T$ is $\hat{\Lambda}_0^T=\left\{ \hat{\Lambda}_t=\mathbf{1}\{t_1\le t \le t_2\}\Lambda_t, \tin \right\}$. Also, for  $\G_t=\sigma(A)\vee\F_t^{\hat{\N}}$, there exists a $(\G_t:\tin)$-predictable process $\Pi_0^T$ such that $\Pi_0^T$ is the $(P,\G_t:\tin)$-intensity of $\hat{N}_0^T$ and $\Pi_t=\E[\hat{\Lambda}_t|\G_t]$   $P$-a.s. for each $\tin$.
\end{Lemma}
\begin{IEEEproof}
See the Appendix. 
\end{IEEEproof}

\section{Channel Model}
The two-user Poisson Channel considered here consists of an encoder $\mathscr{E}_x^T$ and two decoders $\mathscr{D}_y^T$ and $\mathscr{D}_z^T$.
Let $\mathcal{X}_0^T$ denote the set of all waveforms over $[0, T]$ which are non-negative, right continuous with left limits, and peak power limited by unity. This is the set of inputs to the channel, i.e., ${X}_0^T=\{X_t,0\le X_t\le 1,  t\in[0,T]\}$.   The received signal at the first receiver  ${Y}_{0}^T$ is a doubly-stochastic Poisson process with rate $a_yX_0^T+\lambda_y$.
Here $a_y\ge 0$ accounts for possible attenuation of the signal at the first receiver and $\lambda_y\geq0$ is the  dark current intensity due to background noise and is independent of the input process $X_0^T$. 
Similarly the received signal at the second receiver is $Z_0^T$, where 
 ${Z}^T_0$ is a doubly-stochastic Poisson process with rate  $a_zX_0^T+\lambda_z$ with $a_z,\lambda_z \ge 0$.
 
 Let $(\mathcal{X}^T_0,\mathfrak{F}^X)$ denote the input space, where $\mathfrak{F}^X$ is the $\sigma$-field on $\mathcal{X}^T_0$ generated by the open sets of $\mathcal{X}^T_0$ when endowed with the Skorohod topology \cite[Chapter 3, Section 12, p. 121]{Billingsley}. Similarly, let $(\mathcal{N}_0^T,\mathfrak{F}^Y)$ and $(\mathcal{N}_0^T,\mathfrak{F}^Z)$ be the first and second receiver's output space respectively, where $\mathfrak{F}^Y$ and $\mathfrak{F}^Z$ are the  $\sigma$-field   generated by the open sets of $\mathcal{N}^T_0$ when endowed with the Skorohod topology. Let $P_0^{Y_0^T}$ (respectively $P_0^{Z_0^T}$) be the probability measure on the first receiver's (respectively second receiver's) output space such that point process $Y_0^T$ (respectively $Z_0^T$) is a unit-rate Poisson process. Then we will take the output space of the channel to be the product space $(\mathcal{N}_0^T\times\mathcal{N}_0^T,\mathfrak{F}^Y\otimes\mathfrak{F}^Z)$ and our reference measure $P_0$ will be the product measure $P_0=P_0^{Y_0^T}\times P_0^{Z_0^T}$.
 Fix ${x}_0^T\in\mathcal{X}_0^T$, and  let $\Xi_{x_0^T}(\cdot)$ denote the transition probability function from the input space $(\mathcal{X}_0^T,\mathfrak{F}^X)$ to the output space $(\mathcal{N}_0^T\times\mathcal{N}_0^T,\mathfrak{F}^Y\otimes\mathfrak{F}^Z)$.
 The channel is modeled through the following Radon-Nikodym derivative:
\begin{equation}
 \frac{d\Xi_{x_0^T}}{dP_{0}}({y}_0^T,{z}_0^T)=\prod_{u=y,z} p_u(x_0^T,u_0^T),
 \end{equation} 
where
\begin{equation}
\label{eq:channeldef}
p_u(x_0^T,u_0^T)=\exp\left(\int_0^T\log(a_ux_t+\lambda_u)\,d{u}_t+1-(a_ux_t+\lambda_u)\,dt\right),
\end{equation}
where we recall the convention  $\exp(\log(0))=0$. Then due to Girsanov's theorems \cite[Chapter VI, Theorems T2-T4, p. 165-168]{Bremaud}, the process $U_0^T$ has $(\F^{U}_t:\tin)$-intensity $a_ux_0^T+\lambda_u$ under probability measure $\Xi_{x_0^T}$ for $(u,U)\in\{(y,Y),(z,Z)\}$. 
Note that the above model implies that for given $x_0^T\in\mathcal{X}_0^T$,  processes $Y_0^T$ and $Z^T_0$ are independent doubly-stochastic Poisson processes with rate processes $a_yx_0^T+\lambda_y$ and $a_zx_0^T+\lambda_z$ respectively  \cite[Theorem T4, Chapter II, p. 25]{Bremaud}.

Let $M$ be a random variable on a measurable space $(\mathcal{M},\Fk^M)$. For the most part of this paper  $M$ will represent a message intended for  either or both  of the users, in which case $\mathcal{M}$ is a finite set and we will take $\Fk^M$ to be the power set of $\mathcal{M}$. However, in proving Theorem \ref{Thm:LN_MC} to follow, we will take the space $(\mathcal{M},\Fk^M)$ to be isomorphic to the input space $(\mathcal{X}_0^T,\mathfrak{F}^X)$.
Let $\mu_m(dx_0^T)$ denote the transition probability function from $(\mathcal{M},\Fk^M)$ to the input space $(\mathcal{X}_0^T,\mathfrak{F}^X)$. Let $\nu(dm)$ be a probability measure on $(\mathcal{M},\Fk^M)$.
Then these measures induce a joint measure  $P$ on $(\Omega,\F)$, where 
\begin{gather}
\Omega=\mathcal{M}\times\mathcal{X}_0^T\times\mathcal{N}_0^T\times\mathcal{N}_0^T\nn
\F=\Fk^M\otimes\mathfrak{F}^X\otimes\mathfrak{F}^{Y}\otimes\mathfrak{F}^{Z} \nn
P=\nu(dm)\mu_m(dx_0^T) P_0^{Y_0^T}(dy_0^T) P_0^{Z^T_0}(dz_0^T) \prod_{u=y,z} p_u(x_0^T,u_0^T).
\label{EQ:Prob_Space}
\end{gather}

From (\ref{EQ:Prob_Space}), we have $M \rightleftarrows X_0^T \rightleftarrows (Y_0^T,Z_0^T)$ and  $Y_0^T \rightleftarrows X_0^T \rightleftarrows Z_0^T$  forming a Markov chain under $P$. \aaronadd{These Markov chain structures will play a triple role in the upcoming analysis. First, the former implies the finiteness of mutual information quantities  (and hence absolute continuity of measures) of the form $I(A;\U_{t_1}^{t_2})$ for $U\in\{Y,Z\}$, where $A \rightleftarrows X_0^T\rightleftarrows \U_{t_1}^{t_2}$ is a Markov chain   (see Lemma \ref{Le:Abs_cont}). Second, the former allows us compute the log-likelihood ratio martingales through the intensity of the point process  $\U_{t_1}^{t_2}$ (see Theorem \ref{Th:Mu_inf}). Finally, the latter coupling is useful for proving impossibility results (cf.\ Theorem~\ref{Thm:Integral_Identity} to follow). The capacity regions defined subsequently, however, only depend on the two marginal distributions of $Y_0^T$ and $Z_0^T$ given $X_0^T$. Thus our capacity results hold for any channels for which $Y_0^T$ and $Z_0^T$ are Poisson processes with rate $a_y X_0^T + \lambda_y$ and $a_z X_0^T + \lambda_z$, respectively.}

We will assume that the given  filtration $(\F_t:\tin)$, $P$,  and $\F$  satisfy the ``usual conditions"\cite[Chapter III, p. 75]{Bremaud}: $\F$ is  complete with respect to $P$, $\F_t$ is right continuous, and $\F_0$ contains all the $P$-null sets of $\F_t$.

In the rest of this paper we will consider mappings $A$ and $B$  from $\Omega$ in (\ref{EQ:Prob_Space}) to a component space $\mathcal{N}_0^T$ or $\mathcal{M}$ of $\Omega$: 
$A$ can be $M$ itself, or $A$ can be a portion of arrival time process $Y_0^T$ or $Z_0^T$ on the interval $[s_1, s_2]$, which we model as a point process on $\mathcal{N}_0^T$ with no arrival prior to $s_1$ and after $s_2$.
Fix $0\le t_1< t_2\le T$ and consider the process $\U_{t_1}^{t_2}$. Denote by $\hat{U}_t$ its value at time $\tin$. Let $\hat{U}_0^T=\{\hat{U}_t:\tin\}$. Note that $\U_{t_1}^{t_2}$ and $\hat{U}_0^T$ are exactly the same process, but we use $\hat{U}_0^T$ for notational convenience.   We will use the following condition to verify that the mutual information $I(A;\U_{t_1}^{t_2}$) is finite.  

\begin{Lemma}
\label{Le:Abs_cont}
If $A$ is such that $A \rightleftarrows X_0^T\rightleftarrows \U_{t_1}^{t_2}$ forms a Markov chain under measure $P$, then with  $\hat{U}_0^T=\{\hat{U}_t:\tin\}$, where $\hat{U}_t$ is the value of $U_{t_1}^{t_2}$ at time $\tin$
\begin{gather*}
I(A;\U_{t_1}^{t_2}) <\infty,
\end{gather*}
and thus
\begin{gather*}
P^{A,\hat{U}_0^T} \ll P^A\times P^{\hat{U}_0^T} \ll P^A\times P_0^{\hat{U}_0^T},
\end{gather*}
where $P_0^{\hat{U}_0^T}$ is the distribution of process ${\hat{U}_0^T}$ under the measure $P_0^{{U}_0^T}$.
\end{Lemma}
\begin{IEEEproof}
See the Appendix. 
\end{IEEEproof}

In particular the above lemma implies that if $(A,B)\rightleftarrows X_0^T\rightleftarrows \U_{t_1}^{t_2}$ is a Markov chain, then $I(A;\U_{t_1}^{t_2})$ and $I(A;\U_{t_1}^{t_2}|B)$ are finite. The mutual information expressions considered in the sequel will be of this form. The following theorem provides a way of computing such expressions. It will be applied repeatedly in the later sections.

\begin{Theorem}[Log Radon-Nikodym derivatives and Mutual Information Expression]
\label{Th:Mu_inf}
Fix $0\le t_1<t_2\le T$, and  let $(u,U)\in\{(y,Y),(z,Z)\}$.
\begin{enumerate}
\item Log Radon-Nikodym derivatives:\\*
Let $A \rightleftarrows X_0^T\rightleftarrows \U_{t_1}^{t_2}$ be a Markov chain. Denote by $\hat{U}_t$ the  value of $\U_{t_1}^{t_2}$ at time $\tin$. Let $\hat{U}_0^T=\{\hat{U}_t:\tin\}$. Let $\tilde{P}^{A,\hat{U}_0^T}=P^A\times P_0^{\hat{U}_0^T}$. From Lemma \ref{Le:Abs_cont}, ${P}^{A,\hat{U}_{0}^{{T}}}\ll \tilde{P}^{A,\hat{U}_{0}^{{T}}}$. Then
\begin{align}
\label{eq:keyRN}
\log\left(\frac{dP^{A,\hat{U}_0^T}}{d\tilde{P}^{A,\hat{U}_0^T}}\right)=\int_{t_1}^{t_2}\log(a_u\Pi_t+\lambda_u)d{U}_t+1-(a_u\Pi_t+\lambda_u)\,dt,
\end{align}
where the above equality is $P^{A,\hat{U}_0^T}$-a.s., and $\Pi_0^T$ is a $(\sigma(A)\vee\F_t^{\hat{U}},\tin)$-predictable process satisfying for each $t\in [t_1, t_2]$, 
\begin{align*}
\Pi_t=\E[X_t|A,U_{t_1}^t], \quad P^{A,\hat{U}_0^T}\text{-a.s.} \footnotemark
\end{align*}
\footnotetext{Here we have abused notation slightly since this random variable will be defined on a larger probability space in the proof.}%
\vspace{-20pt}
\item Mutual Information Expressions:\\*
Suppose that  the Markov chain $(A,B)\rightleftarrows X_0^T\rightleftarrows \U_{t_1}^{t_2}$ holds. Then 
\begin{align*}
I\left(A;\U_{t_1}^{t_2}\middle|B\right)&=\int_{t_1}^{t_2}\mathbb{E}[\phi_u(\mathbb{E}[X_t|U_{t_1}^t,A,B])]-\mathbb{E}[\phi_u(\mathbb{E}[X_t|U_{t_1}^t,B])]  \,dt \nn
&=\int_{t_1}^{t_2}\mathbb{E}[\phi_u(\mathbb{E}[X_{t-}|U_{t}^{t_2},A,B])]-\mathbb{E}[\phi_u(\mathbb{E}[X_{t-}|U_t^{t_2},B])]  \,dt\nn
&=\int_{t_1}^{t_2}\mathbb{E}[\phi_u(\mathbb{E}[X_t|U^{t_2}_t,A,B])]-\mathbb{E}[\phi_u(\mathbb{E}[X_t|U^{t_2}_t,B])]  \,dt,
\end{align*}
where for $u\in\{y,z\}$ we define
$$
\phi_u(x)=(a_ux+\lambda_u)\log(a_ux+\lambda_u),
$$
with convention that $0\log(0)=0$. Note that
$\phi_u(x)$ is convex and continuous for $x\in[0,1]$.
\end{enumerate}
\end{Theorem}

If $A = X_0^T$, then the identity (\ref{eq:keyRN}) is true by definition
(cf.~(\ref{eq:channeldef})). It is also known
when $A$ is independent of $X_0^T$ 
\cite[(5.6), p.~181]{Bremaud}. Those 
two cases suffice to compute the
quantities $I(X_0^T;Y_0^T)$ and $I(X_0^T;Z_0^T)$. By allowing for arbitrary
$A$ in (\ref{eq:keyRN}), we can compute mutual information expressions 
involving auxiliary random variables, which are needed for multiterminal
problems.


\begin{IEEEproof}
We will consider the measurable space $(\mathcal{A}\times\mathcal{X}_0^T \times\mathcal{N}_0^{T},\Fk^A\otimes\mathfrak{F}^{X}\otimes\mathfrak{F}^{\tilde{U}})$. Here $\mathcal{A}$ is the set on which $A$ takes values  and $\Fk^A$ is its $\sigma$-field.
Let $\tilde{P}^{A,X_0^T,\hat{U}_0^T}$ be defined as
$$
\tilde{P}^{A,X_0^T,\hat{U}_0^T}=P^{A,X_0^T}\times P^{\hat{U}_{0}^{{T}}}_0,
$$
i.e., under $\tilde{P}^{A,X_0^T,\hat{U}_{0}^{T}}$, $\hat{U}_{0}^{T}$ is a Poisson process with  deterministic rate  $\mu_0^T$, independent of $A$ and $X_0^T$, where 
\begin{align*}
\mu_t=\mathbf{1}\{t_1\le t < t_2\}. 
\end{align*}

Let $\G_t=\F_{t}^{\hat{U}}\vee\sigma(A)$. Since under $\tilde{P}^{A,X_0^T,\hat{U}_0^T}$, $A$ is independent of $\hat{U}_{0}^{T}$, using Lemma \ref{Le:MC_Int} we conclude that the $(\tilde{P}^{A,X_0^T,\hat{U}_0^T},\G_t:\tin)$-intensity of $\hat{U}_{0}^{T}$ is $\mu_0^T$.

Since $I(A,X_0^T;\hat{U}_0^T)=I(X_0^T;\hat{U}_0^T)<\infty$, we have that ${P}^{A,X_0^T,\hat{U}_0^T}\ll P^{A,X_0^T}\times P^{\hat{U}_0^{{T}}}$  \cite[Lemma 5.2.3, p. 92]{Gray}. Using the fact that $P^{\hat{U}_0^{{T}}} \ll  P_0^{\hat{U}_0^{{T}}}$ we get \cite[Chapter 1, Exercise 19, p. 22]{Kallenberg}
\begin{align*}
{P}^{A,X_0^T,\hat{U}_0^T} \ll \tilde{P}^{A,X_0^T,\hat{U}_0^T}.
\end{align*}
Let \begin{align*}
\mathcal{L}=\frac{d{P}^{A,X_0^T,\hat{U}_{0}^{T}}}{d\tilde{P}^{A,X_0^T,\hat{U}_{0}^{T}}}
\end{align*}
denote the Radon-Nikodym derivative on the space $(\mathcal{A}\times\mathcal{X}_0^T \times\mathcal{N}_0^{T},\Fk^A\otimes\mathfrak{F}^{X}\otimes\mathfrak{F}^{\tilde{U}})$. 
Consider  the mapping $(a,x_0^T,\hat{u}_0^{T})\to (a,\hat{u}_0^{T})$ from  ($\mathcal{A}\times\mathcal{X}_0^T \times\mathcal{N}_0^{T}$) to ($\mathcal{A} \times\mathcal{N}_0^{T}$).
Since $\sigma(A,U_0^{\hat{T}})=\G_T$,
  $\frac{dP^{A,\hat{U}_0^T}}{d\tilde{P}^{A,\hat{U}_0^T}}$ can be computed as
   \cite[Lemma 5.2.4, p. 96]{Gray}
\begin{align*}
\frac{dP^{A,\hat{U}_0^T}}{d\tilde{P}^{A,\hat{U}_0^T}}=\E_{\tilde{P}}[\mathcal{L}|\G_T].
\end{align*}
Here the subscript $\tilde{P}$ indicates that the expectation is taken with respect to $\tilde{P}^{A,X_0^T,\hat{U}_0^T}$. 
Towards this end define  process $L_0^{T}$ as
\begin{align*}
L_t=\E_{\tilde{P}}[\mathcal{L}|\G_t], \quad t \in [0, T].
\end{align*}
Then $L_0^T$ is a $(\tilde{P}^{A,X_0^T,\hat{U}_0^T},\G_t)$ non-negative absolutely-integrable martingale.

By the martingale representation theorem, the process $L_0^T$ can be written as \cite[Chapter III, Theorem T17, p. 76]{Bremaud} (where we have taken $\sigma(A)$ to be the ``germ $\sigma$-field"):
\begin{align*}
L_t=1+\int_{0}^t K_s(d\hat{{U}}_s- \mu_s ds),
\end{align*}
where $K_0^{T}$ is a $(\G_t:\tin)$-predictable process which satisfies $\int_0^T|K_t|\mu_t\,dt<\infty$ $\tilde{P}^{A,X_0^T,\hat{U}_0^T}$-a.s.
Applying \cite[Lemma 19.5, p. 315]{Liptser}, we can write \newadd{$L_t$} as
\newadd{
\begin{align}
L_t=\exp\left(\int_{0}^t\log(\Psi_s)d\hat{{U}}_s+(1-\Psi_s)\mu_s\,ds\right) \quad \tin,
\label{EQ:L_T}
\end{align}
}
where $\Psi_0^T$ is a non-negative  $(\G_t:\tin)$-predictable process, and $\Psi_t<\infty$ $\tilde{P}^{A,X_0^T,\hat{U}_0^T}$-a.s. for $\tin$. \newadd{Let 
\begin{align*}
\hat{\Psi}_t=\Psi_t\mu_t=\mathbf{1}\{t_1\le t < t_2\}\Psi_t.
\end{align*}
Since the candidate intensity $\hat{\Psi}_0^T$ is not known to satisfy  $\int_0^T\hat{\Psi}_t\,dt<\infty$, we cannot apply \cite[Chapter VI, Theorems T2-T3, p. 166]{Bremaud} directly.
Instead, we first mimic the proof of \cite[Chapter VI, Theorem T3, p. 166]{Bremaud} to get following result.
\begin{Lemma}
\label{le:intensity_proof}
For all non-negative $(\G_t:\tin)$-predictable processes
$C_0^T$
\begin{align*}
\E\left[\int_0^T C_t\hat{\Psi}_t\,dt\right]=\E\left[\int_0^T C_t\,d\hat{U}_t\right],
\end{align*}
where the above expectation is with respect to the measure ${P}^{A,X_0^T,\hat{U}_0^T}$.
\end{Lemma}
\begin{IEEEproof}
See the Appendix.
\end{IEEEproof}
Taking $C_t=1$ in the above equality yields  $$\E\left[\int_0^T \hat{\Psi}_t\,dt\right]=\E\left[\int_0^T \,d\hat{U}_t\right]<\infty.$$ 
 Hence $\int_0^T \hat{\Psi}_t\,dt<\infty$ ${P}^{A,X_0^T,\hat{U}_0^T}$-a.s. and we conclude that the $({P}^{A,X_0^T,\hat{U}_0^T},\G_t:\tin)$-intensity of $\hat{U}_{0}^{T}$ is $\hat{\Psi}_0^{T}$.}

 Moreover  due to uniqueness of predictable intensities \cite[Theorem T12, Chapter II, p. 31]{Bremaud}, from Lemma \ref{Le:MC_Int}, we can take
for $t_1\le t \le t_2$ ${P}^{A,X_0^T,\hat{U}_0^t}\text{-a.s.}$
\begin{align}
{\Psi}_t=a_u\Pi_t+\lambda_u,
\label{EQ:PSi}
\end{align}
where for each $t\in[t_1, t_2]$,
\begin{align}
\Pi_t=\mathbb{E}[{X}_{t}|A,\hat{U}_{0}^{t}].
\label{EQ:PSi1}
\end{align}

Noting that process $\hat{U}_0^T$ has no arrivals prior  to $t_1$ and later than $t_2$, and the same arrivals as $U_0^T$ between $t_1$ and $t_2$, substituting value of $\Psi_t$ from (\ref{EQ:PSi}), (\ref{EQ:L_T}) yields 
\begin{align}
\log\left(\frac{dP^{A,\hat{U}_{0}^{T}}}{d\tilde{P}^{A,\hat{U}_{0}^{T}}}\right)&=\log(L_{T})\nn
&=\int_{t_1}^{t_2}\log(a_u\Pi_t+\lambda_u)d\,{{U}}_t+1-(a_u\Pi_t+\lambda_u)\,dt,
\label{EQ:LLR1}
 \end{align}
 where  $\Pi_t=\mathbb{E}[{X}_{t}|A,{U}_{t_1}^{t}]$ $P^{A,\hat{U}_{0}^{T}}$-a.s. for each $t\in[t_1, t_2]$.
 This proves part (1) of the theorem. 
 
Writing (\ref{EQ:LLR1}) in terms of $\Psi_t$, we get
\begin{align}
\log\left(\frac{dP^{A,\hat{U}_{0}^{T}}}{d\tilde{P}^{A,\hat{U}_{0}^{T}}}\right)=\int_{0}^T\log({\Psi}_t)d\hat{{U}}_t+(1-\Psi_t)\mu_t\,dt,
\label{EQ:LT}
\end{align}
and recalling that   $\Psi_0^{T}$ is $(\G_t:\tin)$-predictable
\begin{align}
\E\left[\log\left(\frac{dP^{A,\hat{U}_{0}^{T}}}{d\tilde{P}^{A,\hat{U}_{0}^{T}}}\right)\right]&=\E\left[\int_{0}^T\log(\Psi_t)d\hat{{U}}_t\right]+\int_0^T (1-\E[\Psi_t])\mu_t\,dt\nn
&=\E\left[\int_{0}^T \log(\Psi_t){\Psi}_t\mu_t\,dt\right]+\int_{t_1}^{t_2} 1-\E[\Psi_t]\,dt\nn
&=\int_{t_1}^{t_2}\E[\Psi_t\log(\Psi_t)]+1-\E[\Psi_t]\,dt \nn
&=\int_{t_1}^{t_2}\E\left[\log(a_u\mathbb{E}[{X}_{t}|A,{U}_{t_1}^{t}]+\lambda_u)(a_u\mathbb{E}[{X}_{t}|A,{U}_{t_1}^{t}]+\lambda_u)\right]+1-(a_u\mathbb{E}[{X}_t]+\lambda_u)\,dt\nn
&=\int_{t_1}^{t_2}\E\left[\phi_u\left(\mathbb{E}[{X}_{t}|A,{U}_{t_1}^{t}]\right)\right]+1-(a_u\mathbb{E}[{X}_t]+\lambda_u)\,dt.
\label{EQ:LLR2}
\end{align}
Similarly 
 \begin{align}
\E\left[\log\left(\frac{dP^{\hat{U}_0^T}}{{dP}_0^{\hat{U}_0^T}}\right)\right]
&=\int_{t_1}^{t_2}\E\left[\phi_u\left(\mathbb{E}[X_t|U_{t_1}^{t}]\right)\right]+1-(a_u\mathbb{E}[X_t]+\lambda_u)\,dt.
\end{align}
Using (\ref{EQ:MI_LLR}) and Lemma \ref{Le:Abs_cont} we can compute the mutual information expression
\begin{align}
I(A;U_{t_1}^{t_2})&=I(A;\hat{U}_{0}^{T})\nn
&=\E\left[\log\left(\frac{dP^{A,\hat{U}_0^T}}{d(P^A\times P^{\hat{U}_0^T})}\right)\right]\nn
&=\E\left[\log\left(\frac{dP^{A,\hat{U}_0^T}\big/ d\tilde{P}^{A,\hat{U}_0^T}}{d(P^A\times P^{\hat{U}_0^T})\big/d\tilde{P}^{A,\hat{U}_0^T}}\right)\right]\nn
&=\E\left[\log\left(\frac{dP^{A,\hat{U}_0^T}\big/ d\tilde{P}^{A,\hat{U}_0^T}}{dP^{\hat{U}_0^T}/dP^{\hat{U}_0^T}_0}\right)\right]\nn
&=\E\left[\log\left(\frac{dP^{A,\hat{U}_{0}^{T}}}{d\tilde{P}^{A,\hat{U}_{0}^{T}}}\right)\right]-\E\left[\log\left(\frac{dP^{\hat{U}_{0}^{T}}}{{dP}_0^{\hat{U}_{0}^{T}}}\right)\right]\nn
&=\int_{t_1}^{t_2}\mathbb{E}[\phi_u(\mathbb{E}[X_t|U_{t_1}^t,A])]-\mathbb{E}[\phi_u(\mathbb{E}[X_t|U_{t_1}^t])]  \,dt.
\end{align}

 Now we  use  Kolmogorov's formula  and the fact that all the mutual information expressions are finite due to Lemma \ref{Le:Abs_cont}:
\begin{align}
I(A;\U_{t_1}^{t_2}|B)=&I(A,B;\U_{t_1}^{t_2})-I(B;\U_{t_1}^{t_2}) \nn
=&\int_{t_1}^{t_2}\mathbb{E}[\phi_u(\mathbb{E}[X_t|U_{t_1}^t,A,B])]-\mathbb{E}[\phi_u(\mathbb{E}[X_t|U_{t_1}^t])]  \,dt \nn
&-\int_{t_1}^{t_2}\mathbb{E}[\phi_u(\mathbb{E}[X_t|U_{t_1}^t,B])]-\mathbb{E}[\phi_u(\mathbb{E}[X_t|U_{t_1}^t])]  \,dt \nn
=&\int_{t_1}^{t_2}\mathbb{E}[\phi_u(\mathbb{E}[X_t|U_{t_1}^t,A,B])]-\mathbb{E}[\phi_u(\mathbb{E}[X_t|U_{t_1}^t,B])]  \,dt.
\end{align}

Now define a new point process $\tilde{U}_{0}^{T}$ as the time-reversed version of the process $\hat{U}_{0}^{T}$. From Lemma \ref{LE:DSPP}, $\tilde{U}_{0}^{T}$ is a doubly-stochastic Poisson process with rate process 
\begin{align*}
\tilde{\Lambda}_{0}^{T}=\{(a_u\tilde{X}_{t}+\lambda_u)\mathbf{1}\{T-t_2\le t < T-t_1\},\tin\},
\end{align*}
where $\tilde{X}_{t}=X_{(T-t)-}$.
Let ${\tilde{U}}_{t}$ denote the value of process $\tilde{U}_{0}^{T}$ . 
Then
\begin{align}
I(A;\U_{t_1}^{t_2}|B)&=I(A;\hat{U}_{0}^{T}|B)\nn
&=I(A;\tilde{U}_{0}^{T}|B)\nn
&=\int_{T-t_2}^{T-t_1}\mathbb{E}[\phi_u(\mathbb{E}[\tilde{X}_s|\tilde{U}_{T-t_2}^{s},A,B])]-\mathbb{E}[\phi_u(\mathbb{E}[\tilde{X}_s|\tilde{U}_{T-t_2}^{s},B])]  \,ds \nn
&=\int_{T-t_2}^{T-t_1}\mathbb{E}[\phi_u(\mathbb{E}[X_{(T-s)-}|U_{T-s}^{t_2},A,B])]-\mathbb{E}[\phi_u(\mathbb{E}[X_{(T-s)-}|U_{T-s}^{t_2},B])]  \,ds \nn
&=\int_{t_1}^{t_2}\mathbb{E}[\phi_u(\mathbb{E}[X_{t-}|U_{t}^{t_2},A,B])]-\mathbb{E}[\phi_u(\mathbb{E}[X_{t-}|U_t^{t_2},B])]  \,dt.
\end{align}
Note that since a c\`adl\`ag process can have at most countably many jumps over a bounded interval $[t_1, t_2]$ \cite[Section 12, Lemma 1, p. 122]{Billingsley}, we have
\begin{align*}
\int_{t_1}^{t_2} \mathbf{1}\{{X}_{t-}\neq X_t\}=0.
\end{align*}
Taking expectation and using Fubini's theorem
\begin{align*}
\frac{1}{t_2-t_1}\int_{t_1}^{t_2} P({X}_{t-}\neq X_t)=0.
\end{align*}
Thus
\begin{align}
P({X}_{S-}= X_S)=1,
\label{EQ:XS}
\end{align}
where we have defined $S$ to be a random variable uniformly distributed over $[t_1, t_2]$ and independent of all other $\sigma$-fields.
We can then write $I(A;\U_{t_1}^{t_2}|B)$ as
\begin{align*}
I(A;\U_{t_1}^{t_2}|B)
&=\int_{t_1}^{t_2}\mathbb{E}[\phi_u(\mathbb{E}[X_{t-}|U_{t}^{t_2},A,B])]-\mathbb{E}[\phi_u(\mathbb{E}[X_{t-}|U_t^{t_2},B])]  \,dt\nn
&=(t_2-t_1)\mathbb{E}[\phi_u(\mathbb{E}[X_{S-}|U_{S}^{t_2},A,B])]-\mathbb{E}[\phi_u(\mathbb{E}[X_{S-}|U_S^{t_2},B])]\nn
&\stackrel{(a)}{=}(t_2-t_1)\mathbb{E}[\phi_u(\mathbb{E}[X_{S}|U_{S}^{t_2},A,B])]-\mathbb{E}[\phi_u(\mathbb{E}[X_{S}|U_S^{t_2},B])]\nn
&\stackrel{}{=}\int_{t_1}^{t_2}\mathbb{E}[\phi_u(\mathbb{E}[X_{t}|U_{t}^{t_2},A,B])]-\mathbb{E}[\phi_u(\mathbb{E}[X_{t}|U_t^{t_2},B])]  \,dt,
\end{align*}
where for (a) we have used (\ref{EQ:XS}). This completes the proof of part (2) of the theorem. 
\end{IEEEproof}
We now derive some  properties of $I(A;U_0^T|B)$.
\begin{Lemma}
\label{Le:Ch_Rule}
If $(A,B)\rightleftarrows X_0^T \rightleftarrows U_0^T$ is a Markov chain, then
\begin{align*}
\lim_{\delta\to 0^+}\frac{1}{\delta}I\left(A;U_t^{t+\delta}\middle| U_{0}^t,B\right)=\mathbb{E}[\phi_u(\mathbb{E}[X_t|U_{0}^t,A,B])]-\mathbb{E}[\phi_u(\mathbb{E}[X_t|U_{0}^t,B])]
\end{align*}
and
\begin{align*}
\lim_{\delta\to 0^+}\frac{1}{\delta}I\left(A;U^t_{t-\delta}\middle| U_t^T,B\right)=\mathbb{E}[\phi_u(\mathbb{E}[X_{t-}|U_t^T,A,B])]-\mathbb{E}[\phi_u(\mathbb{E}[X_{t-}|U_t^T,B])].
\end{align*}
\end{Lemma}
\begin{IEEEproof}
See the Appendix. 
\end{IEEEproof}
\begin{Lemma}
\label{Le:Bounded}
If $A$ and $B$ are such that $(A,B)\rightleftarrows X_0^T \rightleftarrows U_0^T$ is a Markov chain, then both 
$\dfrac{1}{\delta}I\left(A;U_s^{s+\delta}\middle| U_0^s,B\right)$ and $\dfrac{1}{\delta}I\left(A;U^s_{s-\delta}\middle| U_s^T,B\right)$ are bounded uniformly over $s$ and $\delta>0$. 
\end{Lemma}
\begin{IEEEproof}
See the Appendix. 
\end{IEEEproof}

Combining Lemmas \ref{Le:Ch_Rule} and \ref{Le:Bounded} yields the chain rule for mutual information in continuous time.
\begin{Lemma}
\label{Le:Ch_Rule_Cont}
If $(A,B)\rightleftarrows X_0^T \rightleftarrows U_0^T$ is a Markov chain, then
\begin{align*}
I(A;U_{0}^{t}|B)=\lim_{\delta\to 0^+}\frac{1}{\delta}\int_{0}^{t}I\left(A;U_s^{s+\delta}\middle| U_0^s,B\right)\,ds, \nn
I(A;U_{t}^{T}|B)=\lim_{\delta\to 0^+}\frac{1}{\delta}\int_{t}^{T}I\left(A;U_{s-\delta}^{s}\middle| U_s^T,B\right)\,ds.
\end{align*}
\end{Lemma} 
\begin{IEEEproof}
See the Appendix. 
\end{IEEEproof}
We now prove an identity which parallels the Csisz\'{a}r sum identity \cite{Csiszar} for discrete memoryless channels.
\begin{Theorem}
\label{Thm:Integral_Identity}
With the channel model in (\ref{EQ:Prob_Space}):
 \begin{align}
\lim_{\epsilon\to 0^{+}}\int_0^T \frac{1}{\epsilon}I\left(Z_{t-\epsilon}^t;Y_0^t \middle| Z_t^T,M\right)\,dt=\lim_{\epsilon\to 0^{+}}\int_0^T \frac{1}{\epsilon}I\left(Y^{t+\epsilon}_t;Z_t^T \middle| Y_0^t,M\right)\,dt,
\label{EQ:Integral_Identity1}
\end{align}
where we take $U_s^{t_2}=U_0^{t_2}$ if $s< 0$, and $U_{t_1}^s=U_{t_1}^T$ if $s>T$.
This implies
 \begin{align}
&\int_0^T\mathbb{E}[\phi_y(\mathbb{E}[X_t|Y_0^t,M])]-\mathbb{E}[\phi_z(\mathbb{E}[X_t|Z_t^T,M])] \,dt=\nn
&\int_0^T\mathbb{E}[\phi_y(\mathbb{E}[X_t|Y_0^t,Z_t^T,M])]-\mathbb{E}[\phi_z(\mathbb{E}[X_t|Y_0^t,Z_t^T,M])]\,dt.
\label{EQ:Integral_Identity2}
\end{align}
\end{Theorem}
\begin{IEEEproof}
Noting that since $(M,Z_0^T)\rightleftarrows X_0^T \rightleftarrows Y_0^T$ is a Markov chain, the mutual information expressions considered below are finite. Using \cite[Lemma 3.3]{WYNER197851} we get 
\begin{align}
\int_0^T I\left(Z_{t-\epsilon}^t;Y_0^t \middle| Z_t^T,M\right)\,dt&=\int_0^T I\left(Z_{t-\epsilon}^t,Z_t^T;Y_0^t \middle|M\right)-I\left(Z_t^T;Y_0^t \middle|M\right)\,dt\nn
&=\int_0^T I\left(Z_{t-\epsilon}^T;Y_0^t \middle|M\right)\,dt-\int_0^T I\left(Z_t^T;Y_0^t \middle|M\right)\,dt.
\label{EQ:int_id1}
\end{align}
Similarly,
\begin{align}
\int_0^T I\left(Y^{t+\epsilon}_t;Z_t^T \middle| Y_0^t,M\right)\,dt&=\int_0^T I\left(Y^{t+\epsilon}_0;Z_t^T \middle|M\right)\,dt-\int_0^T I\left(Y_0^t;Z_t^T \middle|M\right)\,dt\nn
&=\int_{\epsilon}^{T+\epsilon} I\left(Y^{t}_0;Z_{t-\epsilon}^T \middle|M\right)\,dt-\int_0^T I\left(Y_0^t;Z_t^T \middle|M\right)\,dt.
\label{EQ:int_id2}
\end{align}

From (\ref{EQ:int_id1}) and (\ref{EQ:int_id2}),
\begin{align}
\int_0^T\frac{1}{\epsilon}I\left(Z_{t-\epsilon}^t;Y_0^t \middle| Z_t^T,M\right)\,dt-\int_0^T \frac{1}{\epsilon}I\left(Y^{t+\epsilon}_t;Z_t^T \middle| Y_0^t,M\right)\,dt \nn
=\int_{0}^{T} \frac{1}{\epsilon}I\left(Z_{t-\epsilon}^T;Y^{t}_0 \middle|M\right)\,dt-\int_{\epsilon}^{T+\epsilon} \frac{1}{\epsilon}I\left(Y^{t}_0;Z_{t-\epsilon}^T \middle|M\right)\,dt\nn
=\int_{0}^{\epsilon} \frac{1}{\epsilon}I\left(Y^{t}_0;Z_{t-\epsilon}^T \middle|M\right)\,dt-\int_{T}^{T+\epsilon} \frac{1}{\epsilon}I\left(Y^{t}_0;Z_{t-\epsilon}^T \middle|M\right)\,dt.
\end{align}
Taking limits, we will consider both terms separately
\begin{align}
\lim_{\epsilon\to {0^+}}\int_{0}^{\epsilon} \frac{1}{\epsilon}I\left(Y^{t}_0;Z_{t-\epsilon}^T \middle|M\right)\,dt &\stackrel{(a)}{\le} \lim_{\epsilon\to {0^+}}\int_{0}^{\epsilon} \frac{1}{\epsilon}I\left(Y^{t}_0;Z_{0}^T \middle|M\right)\,dt \nn
&\stackrel{(b)}{\le} \lim_{\epsilon\to {0^+}}\int_{0}^{\epsilon}\frac{1}{\epsilon} I\left(Y^{\epsilon}_0;Z_{0}^T \middle|M\right)\,dt \nn
&=\lim_{\epsilon\to {0^+}}I\left(Y^{\epsilon}_0;Z_{0}^T \middle|M\right) \nn
&\stackrel{(c)}{=}0,
\end{align}
where, for (a) and (b) we have used the fact that $I(\U_{t_1}^{t_2};A|B)$ is monotonic in $t_1$ and $t_2$ since
\begin{equation*}
I\left(A;\U_{t_1}^{t_2}\middle|B\right)=\int_{t_1}^{t_2}\mathbb{E}[\phi_u(\mathbb{E}[X_t|U_{t_1}^t,A,B])]-\mathbb{E}[\phi_u(\mathbb{E}[X_t|U_{t_1}^t,B])]  \,dt.
\end{equation*}
As the integrand is non-negative due to Jensen's inequality, $I\left(A;\U_{t_1}^{t_2}\middle|B\right)$ is non-increasing in $t_1$ for fixed $t_2$ and non-decreasing in $t_2$ for fixed $t_1$.
Also, since the integrand is bounded,
\begin{align*}
\lim_{t_2 \to t_1^{+}}I\left(A;\U_{t_1}^{t_2}\middle|B\right)=0.
\end{align*}
This gives (c).
Similarly,
\begin{align*}
\lim_{\epsilon\to {0^+}}\frac{1}{\epsilon}\int_{T}^{T+\epsilon} I\left(Y^{t}_0;Z_{t-\epsilon}^T \middle|M\right)\,dt=0.
\end{align*}
This proves part (1).
Since $\frac{1}{\epsilon}I\left(Z_{t-\epsilon}^t;Y_0^t \middle| Z_t^T,M\right)$ and $\frac{1}{\epsilon}I\left(Y^{t+\epsilon}_t;Z_t^T \middle| Y_0^t,M\right)$ are bounded over $\epsilon>0$ from Lemma \ref{Le:Bounded}, we use the dominated convergence theorem to swap the integral and limit in (\ref{EQ:Integral_Identity1}) to get
\begin{align}
\int_0^T \lim_{\epsilon\to 0^{+}}\frac{1}{\epsilon}I\left(Z_{t-\epsilon}^t;Y_0^t \middle| Z_t^T,M\right)\,dt=\int_0^T \lim_{\epsilon\to 0^{+}}\frac{1}{\epsilon}I\left(Y^{t+\epsilon}_t;Z_t^T \middle| Y_0^t,M\right)\,dt.
\label{EQ:Integral_Identity5}
\end{align}

Taking $U=Z$, $A=Y_0^t$ and $B=M$ in the left-hand side of (\ref{EQ:Integral_Identity5}), Lemma \ref{Le:Ch_Rule} gives
\begin{align*}
\int_0^T \lim_{\epsilon\to 0^{+}}\frac{1}{\epsilon}I\left(Z_{t-\epsilon}^t;Y_0^t \middle| Z_t^T,M\right)\,dt=\int_0^T  \mathbb{E}[\phi_z(\mathbb{E}[X_{t-}|Y_0^t,Z_t^T,M])]-\mathbb{E}[\phi_z(\mathbb{E}[X_{t-}|Z_t^T,M])]\,dt.
\end{align*}
Since $X_0^T$ is a c\`adl\`ag process, we can repeat the same argument as in the proof of Theorem \ref{Th:Mu_inf} to replace $X_{t-}$ in the above integral with $X_{t}$.  We get 
\begin{align}
\int_0^T \lim_{\epsilon\to 0^{+}}\frac{1}{\epsilon}I\left(Z_{t-\epsilon}^t;Y_0^t \middle| Z_t^T,M\right)\,dt=\int_0^T  \mathbb{E}[\phi_z(\mathbb{E}[X_{t}|Y_0^t,Z_t^T,M])]-\mathbb{E}[\phi_z(\mathbb{E}[X_{t}|Z_t^T,M])]\,dt.
\label{EQ:Integral_Identity3}
\end{align}
Similarly, taking $U=Y$, $A=Z_t^T$ and $B=M$ in the right hand side of (\ref{EQ:Integral_Identity5}), Lemma \ref{Le:Ch_Rule} gives
\begin{align}
\int_0^T \lim_{\epsilon\to 0^{+}}\frac{1}{\epsilon}I\left(Y^{t+\epsilon}_t;Z_t^T \middle| Y_0^t,M\right)\,dt=\int_0^T\mathbb{E}[\phi_y(\mathbb{E}[X_t|Y_0^t,Z_t^T,M])]-\mathbb{E}[\phi_y(\mathbb{E}[X_t|Y_0^t,M])]\,dt.
\label{EQ:Integral_Identity4}
\end{align}  
The second part of the lemma now follows since (\ref{EQ:Integral_Identity3}) and (\ref{EQ:Integral_Identity4}) are equal from (\ref{EQ:Integral_Identity5}). 
\end{IEEEproof}

\newadd{\section{Comparison of Two Receivers}}
Motivated by the definition for the discrete memoryless channels \cite{Korner}, we define a less noisy receiver and a more capable receiver  for the two-user Poisson channel as follows.  
\begin{Def}[Less Noisy Receiver]
Receiver 1 is said to be \textit{less noisy} than receiver 2 if $I(M;Y_0^T)\ge I(M;Z_0^T)$ for all possible $M$ in (\ref{EQ:Prob_Space}), where $M\rightleftarrows X_0^T\rightleftarrows(Y_0^T,Z_0^T)$ is a Markov chain.
\end{Def}

\begin{Def}[More Capable Receiver]
Receiver 1 is said to be \textit{more capable} than receiver 2 if $I(X_0^T;Y_0^T)\ge I(X_0^T;Z_0^T)$ for all probability measures on the input space $(\mathcal{X}_0^T,\Fk^X)$.
\end{Def}
We shall call a channel with \newadd{a} less  noisy receiver to be a less noisy Poisson channel and similarly a channel with \newadd{a} more capable receiver to be a more capable Poisson channel.
\begin{Theorem}
\label{Thm:LN_MC}
In a two-user Poisson channel the following conditions are equivalent:
\begin{enumerate}[(I)]
\item $\Phi(x)=\phi_y(x)-\phi_z(x)$ is a convex function over $[0, 1]$.
\item Receiver 1 is less noisy than receiver 2.
\item Receiver 1 is more capable than receiver 2.
\newadd{\item The channel parameters satisfy 
\begin{itemize}
\item $a_y\ge a_z$ and $a_y^2\lambda_z\ge a_z^2\lambda_y$; or
\item  $0< a_y< a_z$ and  $a_y^2(a_z+\lambda_z)\ge a_z^2(a_y+\lambda_y)$.
\end{itemize}}
\end{enumerate}

  \end{Theorem}
\begin{IEEEproof}
To prove (I) implies (II), note that Theorem \ref{Th:Mu_inf} yields
\setlength{\arraycolsep}{0.0em}
\begin{eqnarray}
I(M;Y_0^T)-I(M;Z_0^T)&{}={}&\int_0^T\mathbb{E}[\phi_y(\mathbb{E}[X_t|Y_0^t,M])]- \mathbb{E}[\phi_y(\mathbb{E}[X_t|Y_0^t])]\,dt\nonumber\\
&&{-}\:\int_0^T\mathbb{E}[\phi_z(\mathbb{E}[X_t|Z_t^T,M])]-\mathbb{E}[\phi_z(\mathbb{E}[X_t|Z_t^T])]\,dt\\
&{}={}&\int_0^T\mathbb{E}[\phi_y(\mathbb{E}[X_t|Y_0^t,M])]-\mathbb{E}[\phi_z(\mathbb{E}[X_t|M,Z_t^T])]\nonumber\,dt\\
&&{-}\:\int_0^T\mathbb{E}[\phi_y(\mathbb{E}[X_t|Y_0^t])]-\mathbb{E}[\phi_z(\mathbb{E}[X_t|Z_t^T])]\nonumber\,dt \\
&\stackrel{(a)}{=}&\int_0^T\mathbb{E}[\phi_y(\mathbb{E}[X_t|Y_0^t,Z_t^T,M])]-\mathbb{E}[\phi_z(\mathbb{E}[X_t|Y_0^T,Z_t^T,M])]\nonumber\,dt\\
&&{-}\:\int_0^T\mathbb{E}[\phi_y(\mathbb{E}[X_t|Y_0^t,Z_t^T])]-\mathbb{E}[\phi_z(\mathbb{E}[X_t|Y_0^t,Z_t^T])]\nonumber\,dt\\
&{}={}&\int_0^T\mathbb{E}[\Phi(\mathbb{E}[X_t|Y_0^t,Z_t^T,M])]- \mathbb{E}[\Phi(\mathbb{E}[X_t|Y_0^t,Z_t^T])]\,dt,
\end{eqnarray}
\setlength{\arraycolsep}{0.0em}
where (a) is due to Theorem \ref{Thm:Integral_Identity}.  Since $\Phi(x)$ is a convex function, Jensen's inequality gives
\setlength{\arraycolsep}{0.0em}
\begin{eqnarray}
I(M;Y_0^T)-I(M;Z_0^T)&{}={}&\int_0^T\mathbb{E}[\Phi(\mathbb{E}[X_t|Y_0^t,Z_t^T,M])]- \mathbb{E}[\Phi(\mathbb{E}[X_t|Y_0^t,Z_t^T])]\,dt\nonumber\\
&{}\ge{}&0.
\end{eqnarray}
\setlength{\arraycolsep}{0.0em}
Note that (II) implies (III) trivially.
We now prove that (III) implies (I).
There exists a sequence of input distributions (indexed by $n$), such that $X_0^T$ is binary and stationary with the following limit\cite{Kabanov,Davis}
\begin{align*}
\lim_{n\to\infty}\E\left[\phi_u\left(\E[X_t|U_0^t]\right)\right]=\phi_u(\E[X_t]).
\end{align*}
Thus choosing $X_t$ such that $P(X_t=p)=1-P(X_t=q)=\alpha$,  $0\le\alpha \le 1$ and taking the limit gives
\begin{align*}
\alpha\phi_y(p)+(1-\alpha)\phi_y(q)-\phi_y(\alpha p+(1-\alpha)q)\ge\alpha\phi_z(p)+(1-\alpha)\phi_z(q)-\phi_z(\alpha p+(1-\alpha)q).
\end{align*}
Therefore 
\begin{align*}
\alpha\Phi(p)+(1-\alpha)\Phi(q)\ge\Phi(\alpha p+(1-\alpha)q).
\end{align*}
Hence $\Phi(x)$ is a convex function.

The channel parameters for which the channel is less noisy can be obtained by calculating conditions under which the second derivative of $\Phi(x)$ is non-negative for $0\le x\le 1$.
\end{IEEEproof}

Note that these channel parameters include the parameters for which the channel is known to be stochastically degraded \cite{Lapidoth} 
\begin{align}
\label{EQ:Stoc_Deg}
a_y\ge a_z, \qquad   a_y\lambda_z\ge a_z \lambda_y.
\end{align}

The conditions given in Theorem~\ref{Thm:LN_MC} differ from the conditions
under which the discretized Poisson channel is more capable.
A discretized Poisson channel is a discrete memoryless channel in which  the input is binary and constant over $\tau$-duration intervals, where $\tau$ is very small. The output in an interval is taken to be $``1"$ if there are one or more arrivals during this interval and $``0"$ otherwise. Wyner\cite{WYNER197851} shows that, for the purposes of
reliable communication, the Poisson channel is equivalent \newadd{to} its
discretized version, so that coding theorems for the former may
be inferred from the latter. This equivalence carries over to
Poisson broadcast channels~\cite{Lapidoth}.

Kim \textit{et al.}~\cite{Kim} determine the range of parameters 
under which the discretized Poisson broadcast channel is less 
noisy and more capable. The conditions under which the discretized
channel is less noisy match those in Theorem~\ref{Thm:LN_MC}. The conditions
for the discretized channel to be more capable, however, are 
strictly weaker: if $a_y=0.4$, \newadd{$\lambda_y=0$}, $a_z=\lambda_z=1$, 
for example, the discretized channel is  more capable \cite[Theorem 1]{Kim}, 
whereas the continuous-time, continuous-space channel considered here is 
not. \newadd{To see the reason behind this, consider a sequence of input distributions (indexed by $n$) as in the proof of Theorem~\ref{Thm:LN_MC},  such that $X_0^T$ is binary and stationary with the following limit for $u\in\{y,z\}$~\cite{Kabanov,Davis}
\begin{align*}
\lim_{n\to\infty}\E\left[\phi_u\left(\E[X_t|U_0^t]\right)\right]=\phi_u(\E[X_t]).
\end{align*}
Then choosing $X_t$ such that $P(X_t=1)=P(X_t=0.9)=0.5$,  and taking the limit gives
\begin{align*}
\lim_{n\to\infty}\frac{1}{T}I(X_0^T;Z_0^T)\approx 6.41 \times 10^{-4} >5.26\times 10^{-4}\approx
\lim_{n\to\infty}\frac{1}{T}I(X_0^T;Y_0^T).
\end{align*}
If $X_0^T$ only takes values in $\{0, 1\}$, on the other hand,
then this inequality is impossible.} Of course, for the purposes of
reliable communication, $X_0^T$ need only \aaronadd{take} values in $\{0, 1\}$,
as noted above. 

Nair~\cite{Nair} defines one discrete memoryless channel to be 
\emph{essentially more capable} than another if a condition
similar to the usual definition of ``more capable'' holds under
a restricted set of input distributions that dominates all others
in certain single-letter mutual information expressions.
The statement that one discretized Poisson channel is
more capable than another thus translates into something
akin to ``essentially more capable'' when expressed in terms of the 
underlying continuous Poisson channels. This analogy is not exact,
however, in that ``essentially more capable'' is defined in terms of 
mutual information expressions while the reduction from the Poisson
channel to its discretized version is operational. 
All of this indicates that some care is required
when translating statements between the Poisson channel and its
discretized version.

We next apply the results obtained thus far to characterize
the capacity (regions) for several multi-receiver communication
problems. The first of these is the more-capable Poisson
broadcast channel. Our result here is less general than that
obtained by Kim \emph{et al.}~\cite{Kim}, although
our proof is more self contained in that it does not require
a discretization argument. We then prove new results on the Poisson
broadcast channel with degraded message sets and the Poisson
wiretap channel. 
 
\section{More Capable Poisson Broadcast Channel}

We first prove several lemmas. Let $T_n=n\tau$ for some  $\tau>0$.
Construct an auxiliary process $V_0^{T_n}$  to be piecewise constant, taking value in the finite alphabet $\mathcal{V}=\{1,\dots,K_v\}$  as follows. We divide the interval $[0, {T_n}]$ into $n$ intervals each of equal length $\tau$. The process will be constant on each of these sub-intervals with value given by 
\begin{align}
V_t=\bar{V}_i \mbox{ for } (i-1)\tau \le t < i\tau, \quad i=1,2,\dots,n
\label{EQ:Vt}
\end{align}
where $\bar{V}_i$'s are independent and identically \newadd{distributed} random variables with $P(\bar{V}_i=j)=\alpha_j$, $j\in\mathcal{V}$.
Let $\mathcal{V}_0^{T_n}$ denote the collection of all such processes. The input waveform $X_0^{T_n}$ is binary and piecewise constant with  
\begin{equation}
X_t=\bar{X}_i \mbox{ for } (i-1)\tau \le t < i\tau, \quad i=1,2,\dots,n
\label{EQ:Xt}
\end{equation}
where
\begin{equation}
P(\bar{X}_i=1|\bar{V}_i=j)=1-P(\bar{X}_i=0|\bar{V}_i=j)=p_j.
\label{EQ:barX}
\end{equation}
The following lemma shows that with the above input to the channel, we have essentially decomposed the single channel use into $n$ independent and identical channel uses.
\begin{Lemma}
\label{Le:Ind_rate}
Let $\U_t^{(i)}$ be the point process corresponding to the arrival time process $U_{(i-1)\tau}^{i\tau}$. The joint distribution of processes $(\bar{V}_i,\bar{X}_i,\U_t^{(i)}:t\in [(i-1)\tau,i\tau])$ is independent and identical across the disjoint blocks for $i=1,\dots,n$ and $U\in\{Y,Z\}$. 
\end{Lemma}

For fixed $V_0^{T_n}\in\mathcal{V}_0^{T_n}$, let $P^{X_0^{T_n}|V_0^{T_n}}$ denote the probability measure on the input space from the construction in (\ref{EQ:Vt})-(\ref{EQ:barX}). Then the probability measure on $(\mathcal{N}_0^{T_n},\Fk^Y)$ for fixed $V_0^{T_n}$  is \cite[Lemma 1.41, p. 21]{Kallenberg}
\begin{align*}
P^{Y_0^{T_n}|V_0^{T_n}}(dy_0^{T_n})=\int_{\mathcal{X}_0^T}P^{X_0^{T_n}|V_0^{T_n}}p_y(x_0^{T_n},y_0^{T_n})P_0(dy_0^{T_n}).
\end{align*}
Let \begin{align}
{Q}^{V_0^{T_n},X_0^{T_n},Y_{0}^{T_n}}=P^{V_0^{T_n}}\times P_{}^{X_0^{T_n}|V_0^{T_n}} \times P_{}^{Y_{0}^{T_n}|V_0^{T_n}}. 
\label{EQ:Measure_Q}
\end{align}
Hence under ${Q}^{V_0^{T_n},X_0^{T_n},Y_{0}^{T_n}}$, the joint distribution of $(V_0^{T_n},X_0^{T_n})$ and $ (V_0^{T_n},Y_0^{T_n})$ is the same as that under $P$, and $ X_0^{T_n}\rightleftarrows V_0^{T_n} \rightleftarrows Y_0^{T_n}$ forms a Markov chain.
\begin{Def}
\label{Def:MID}
The following mutual information densities are defined whenever the corresponding Radon-Nikodym derivatives exist and are strictly positive, in which case we will say that the mutual information densities exist.
\begin{align*}
\mathfrak{i}(X_0^{T_n};Y_0^{T_n})&=\log\left(\frac{dP^{X_0^{T_n},Y_0^{T_n}}}{d(P^{X_0^{T_n}}\times P^{Y_0^{T_n}})}\right)\nn
\ii(X_0^{T_n};Y_0^{T_n}|V_0^{T_n})&=\log\left(\frac{d{P}^{V_0^{T_n},X_0^{T_n},Y_{0}^{T_n}}}{d{Q}^{V_0^{T_n},X_0^{T_n},Y_{0}^{T_n}}}\right)\nn
\mathfrak{i}(V_0^{T_n};Z_0^{T_n})&=\log\left(\frac{dP^{V_0^{T_n},Z_0^{T_n}}}{d(P^{V_0^{T_n}}\times P^{Z_0^{T_n}})}\right).
\end{align*}
\end{Def}
\begin{Lemma}
\label{Le:Conv}
The mutual information densities in Definition \ref{Def:MID} exist, and  for all $\epsilon>0$ there exists $\bar{\tau}$ and $N$ such that if $n\ge N$ and $\tau \le \bar{\tau}$ then
\begin{align}
P\left(\left|\frac{1}{T_n}\mathfrak{i}(X_0^{T_n};Y_0^{T_n})-\left(\E\left[\phi_y(X_0)\right]-\phi_y(\E[X_0])\right)\right|>\epsilon\right) & \le\epsilon\nn
P\left(\left|\frac{1}{T_n}\mathfrak{i}(V_0^{T_n};Z_0^{T_n})-\left(\E\left[\phi_z\left(\E[X_0|\bar{V}_1]\right)\right]-\phi_z\newadd{(}\E[X_0])\right)\right| >\epsilon\right) & \le\epsilon\nn
P\left(\left|\frac{1}{{T_n}}\mathfrak{i}(X_0^{T_n};Y_0^{T_n}|V_0^{T_n})-\left(\E\left[\phi_y(X_0)\right]-\E\left[\phi_y\left(\E[X_0|\bar{V}_1]\right)\right]\right)\right|>\epsilon\right)& \le\epsilon.
\end{align}
\end{Lemma}
\begin{IEEEproof}
See the Appendix. 
\end{IEEEproof}
\begin{Lemma}
\label{Le:ln_inq}
If user 1 is more capable than user 2, then
\begin{align}
\int_0^T\mathbb{E}[\phi_z(\mathbb{E}[X_t|M,Y_0^t])] \,dt \geq \int_0^T\mathbb{E}[\phi_z(\mathbb{E}[X_t|M,Z_t^T])] \,dt.
\end{align}
\end{Lemma}
\begin{IEEEproof}
See the Appendix. 
\end{IEEEproof}
\subsection{Encoding and Decoding}
An \emph{$(L_y,L_z,T)$ code} for the Poisson broadcast channel consists of a source (equipped with an encoder $\mathscr{E}_x^T$) and two receivers each with a decoder ($\mathscr{D}_{y}^T$ and $\mathscr{D}_{z}^T$).  The source has two independent messages $M_y$ and $M_z$ for the first and second user, respectively, where $M_y$ and $M_z$ are uniformly distributed on sets $\mathcal{M}_y=\{1,\dots, L_y\}$ and $\mathcal{M}_z=\{1,\dots, L_z\}$, respectively. 

Given messages $M_y$ and $M_z$ the encoder selects a waveform in $\mathcal{X}_0^T$
\begin{equation}
\mathscr{E}_x^T: \mathcal{M}_y \times \mathcal{M}_z \rightarrow \mathcal{X}_0^T. 
\end{equation} 
Let $\Delta_{x_0^T}(dx_0^T)$ be the Dirac measure on the input space  induced by the given messages $m_y$, $m_z$, and the encoder $\mathscr{E}_x^T$. 
Then  the probability space $(\Omega,\F, P)$ is  
\begin{gather}
\Omega=\mathcal{M}_y\times\mathcal{M}_z\times\mathcal{X}_0^T\times\mathcal{N}_0^T\times\mathcal{N}_0^T\nn
\F=2^{\mathcal{M}_y\times\mathcal{M}_z}\otimes\Fk^X\otimes\Fk^{Y}\otimes\Fk^{Z}\nn
P=\nu(m_y,m_z)\Delta _{\mathscr{E}_x^T(m_y,m_z)}(dx_0^T)P_0^Y(dy_0^T) P_0^Z(dz_0^T) \prod_{u=y,z} p_u(x_0^T,u_0^T).
\label{EQ:Prob_space_bc}
\end{gather}
Here $\nu(m_y,m_z)$ is the uniform distribution on $\mathcal{M}_y\times\mathcal{M}_z$, and $2^{\mathcal{M}_y\times\mathcal{M}_z}$ is the power set of $\mathcal{M}_y\times\mathcal{M}_z$.

On observing $Y_0^T$ and $Z_0^T$,  each decoder chooses a message 
\begin{align}
\mathscr{D}_{y}^T : \mathcal{N}_0^T\rightarrow\mathcal{M}_y \nn
\mathscr{D}_{z}^T : \mathcal{N}_0^T\rightarrow\mathcal{M}_z.
\end{align}
The average probability of error for this code is 
\begin{equation}
\text{P}_e=\frac{1}{L_yL_z}\sum_{m_y=1,m_z=1}^{L_y,L_z}P\left\{\{\mathscr{D}^T_{y}(Y_0^T)\neq m_y\} \bigcup \{\mathscr{D}^T_{z}(Z_0^T)\neq m_z\} \middle| M_y=m_y,M_z=m_z \right\}.
\end{equation}
A rate pair $(R_y,R_z)$ is said to be \emph{achievable} if for all $\epsilon>0$ and sufficiently large $T$, there exists an $(L_y,L_z,T)$ code such that 
\begin{align}
\frac{\log(L_y)}{T}&\ge R_y-\epsilon \nn
\frac{\log(L_z)}{T}&\ge R_z-\epsilon \nn
\text{P}_e&\le\epsilon.
\label{EQ:BC_achv}
\end{align}
The capacity region $(C_y,C_z)$ is the closure of  achievable rate pairs. 

\begin{Theorem}[Capacity of more capable  Poisson broadcast channel]
\label{Th:Cap_LN}
The capacity of the more capable  Poisson broadcast channel when receiver 1 is more capable than receiver 2 is given by the convex hull of the union over all $0\le\alpha\le\frac{1}{2}$ and $0\le p,q \le 1$ of rate pairs  satisfying 
\begin{align*}
 R_y\le C_y&=\alpha(p\phi_y(1)+(1-p)\phi_y(0)-\phi_y(p))+(1-\alpha)(q\phi_y(1)+(1-q)\phi_y(0)-\phi_y(q)) \\
 R_z\le C_z&=\alpha\phi_z(p)+(1-\alpha)\phi_z(q)-\phi_z(\alpha p+(1-\alpha)q).
\end{align*}
\end{Theorem}
Although the proof of the above theorem can be found in \cite{Kim}, we provide an alternate proof using tools derived from stochastic calculus without resorting to the discretization of the continuous-time, continuous-space Poisson channel. Similar proof techniques will be used in proving the capacity theorem of the Poisson broadcast channel with degraded message set to follow. The achievability and converse arguments are provided in next two subsections. 
\subsection{Achievability}
We first note that that $C_y$ and $C_z$ are  upper bounded by the point-to-point capacity of the single-receiver Poisson channel to the first and second user respectively, which for the channel parameters $(a_u,\lambda_u)$, $u\in\newadd{\{y,z\}}$ is given by \cite{Kabanov,Wyner,Davis}
\begin{align*}
C_u^{\text{pp}}=\max_{0\le \kappa \le 1}\kappa\phi_u(1)+(1-\kappa)\phi_u(0)-\phi_u(\kappa).
\end{align*} 
Let $\kappa=\alpha p +(1-\alpha)q$, and using the convexity of $\phi_u$ :
\begin{align*}
C_y&=\alpha(p\phi_y(1)+(1-p)\phi_y(0)-\phi_y(p))+(1-\alpha)(q\phi_y(1)+(1-q)\phi_y(0)-\phi_y(q))\nn
&=(\alpha p+(1-\alpha)q)\phi_y(1)+(\alpha(1-p)+(1-\alpha)(1-q)\phi_y(0))-(\alpha\phi_y(p)+(1-\alpha)\phi_y(q))\nn
&\le (\alpha p+(1-\alpha)q)\phi_y(1)+(\alpha(1-p)+(1-\alpha)(1-q)\phi_y(0))-\phi_y(\alpha p +(1-\alpha)q)\nn
&=\kappa\phi_y(1)+(1-\kappa)\phi_y(0)-\phi_y(\kappa)\nn
&\le C_y^{\text{pp}}.
\end{align*}
Likewise
\begin{align*}
C_z&=\alpha\phi_z(p)+(1-\alpha)\phi_z(q)-\phi_z(\alpha p+(1-\alpha)q)\nn
&\le \alpha p\phi_z(1)+\alpha (1-p)\phi_z(0)+(1-\alpha)q\phi_z(1)+(1-\alpha)(1-q)\phi_z(0)-\phi_z(\alpha p+(1-\alpha)q)\nn
&=\kappa\phi_z(1)+(1-\kappa)\phi_z(0)-\phi_z(\kappa)\nn
&\le C_z^{\text{pp}}.
\end{align*}
Thus if $\alpha$, $p$, and $q$ are such that either $C_y$ or $C_z$ is zero, then achievability follows from the point-to-point achievability argument in \cite{Wyner}. Hence we consider the cases when both of these quantities are strictly positive.
Let $T_n=n\tau$ for some finite $\tau>0$.
Construct an auxiliary process $V_0^{T_n}$  to be a piecewise constant binary-valued process.  We divide the interval $[0, {T_n}]$ into $n$ intervals each of equal length $\tau$. The process will be constant on each of these sub-intervals with value given by 
\begin{align}
V_t=\bar{V}_i \mbox{ for } (i-1)\tau \le t < i\tau, \quad i=1,2,\dots,n
\label{EQ:1Vt}
\end{align}
where $\bar{V}_i$'s are independent and identically \newadd{distributed} Bernoulli random variables with $P(\bar{V}_i=1)=\alpha$. 

The input waveform $X_0^{T_n}$ is binary and piecewise constant with  
\begin{equation}
X_t=\bar{X}_i \mbox{ for } (i-1)\tau \le t < i\tau, \quad i=1,2,\dots,n
\label{EQ:1Xt}
\end{equation}
where
\begin{align}
P(\bar{X}_i=1|\bar{V}_i=1)=1-P(\bar{X}_i=0|\bar{V}_i=1)=p\nn
P(\bar{X}_i=1|\bar{V}_i=0)=1-P(\bar{X}_i=0|\bar{V}_i=0)=q.
\label{EQ:1barX}
\end{align}
An application of Lemma \ref{Le:Conv} yields:
\begin{Lemma}
\label{Le:Conv1}
Let  $\tilde{C}_y=\alpha\phi_y(p)+(1-\alpha)\phi_y(q)-\phi_y(\alpha p+(1-\alpha)q)$. For all $\epsilon>0$ there exist $\bar{\tau}$ and $N$ such that if $n\ge N$ and $\tau \le \bar{\tau}$, then
\begin{align*}
P\left(\left|\frac{1}{T_n}\mathfrak{i}(X_0^{T_n};Y_0^{T_n})- (C_y+\tilde{C}_y)\right|>\epsilon\right)\le\epsilon\nn
P\left(\left|\frac{1}{T_n}\mathfrak{i}(V_0^{T_n};Z_0^{T_n})-C_z\right|>\epsilon\right)\le\epsilon\nn
P\left(\left|\frac{1}{{T_n}}\mathfrak{i}(X_0^{T_n};Y_0^{T_n}|V_0^{T_n})-C_y\right|>\epsilon\right)\le\epsilon.
\end{align*}
\end{Lemma}

\newadd{
\begin{IEEEproof}
See the Appendix. 
\end{IEEEproof}}

\subsubsection{Encoding Operation}
We use superposition coding. Fix  $\delta>0$, and let $R_y=C_y-\delta$ and $R_z=C_z-\delta$. We generate $L_z=\exp(T_nR_z)$ many $V_0^{T_n}$ waveforms (indexed by $j=1,\dots, L_z $) independently according to (\ref{EQ:1Vt}). For  each $V_0^{T_n}(j)$,  we generate $L_y=\exp(T_nR_y)$ many independent $X_0^{T_n}$ waveforms (indexed by $i=1,\dots,L_y$) according to (\ref{EQ:1Xt}) and (\ref{EQ:1barX}). To transmit messages $(M_y,M_z)$, encoder sends $X_0^{T_n}(M_y,M_z)$ over the channel.  
\subsubsection{Decoding Operation}
For a received $Z_0^{T_n}$, the second receiver considers only those  $V_0^{T_n}$ for which  both  $\frac{1}{T_n}\log\left(\frac{dP^{V_0^{T_n},Z_{0}^{T_n}}}{d\tilde{P}^{V_0^{T_n},Z_{0}^{T_n}}}\right)$ and  $\frac{1}{T_n}\log\left(\frac{dP^{Z_{0}^{T_n}}}{d\tilde{P}^{Z_{0}^{T_n}}}\right)$ (calculated using Theorem \ref{Th:Mu_inf}) are finite. We note that $\{\Pi_t:\tin\}$ as in Theorem~\ref{Th:Mu_inf} is $V_0^{T_n}$, $Z_0^{T_n}$ measurable.  
 It  seeks the unique $j$ among all such waveforms  such that 
\begin{align}
\frac{1}{{T_n}}\ii(V_0^{T_n}(j);Z_0^{T_n})= \frac{1}{T_n}\log\left(\frac{dP^{V_0^{T_n},Z_{0}^{T_n}}}{d\tilde{P}^{V_0^{T_n},Z_{0}^{T_n}}}\right)-\frac{1}{T_n}\log\left(\frac{dP^{Z_{0}^{T_n}}}{d{P}_0^{Z_{0}^{T_n}}}\right)\ge C_z-\gamma_z
\label{EQ:Dec_z_LN}
\end{align}
for some $\gamma_z>0$, and outputs $\hat{M}_z=j$. If the decoder does not find any such $V_0^{T_n}$, or if it finds more than  one $V_0^{T_n}$ that satisfy (\ref{EQ:Dec_z_LN}), then the decoder arbitrarily outputs some $\hat{M}_z\in [1, \dots, L_z]$. 

The first receiver decodes both $M_y$ and $M_z$, and we declare an error if either or both messages are decoded incorrectly. It seeks a unique $i$ and $j$ that satisfy both
\begin{align}
\frac{1}{{T_n}}\ii(X_0^{T_n}(i,j);Y_0^{T_n})\ge {C}_y+\tilde{C}_y-\gamma_y
\label{EQ:Dec_y11}
\end{align}
and
\begin{align}
\frac{1}{{T_n}}\ii(X_0^{T_n}(i,j);Y_0^{T_n}|V_0^{T_n}(j))\ge C_y-\gamma_y.
\label{EQ:Dec_y21}
\end{align}
The decoder considers only those $X_0^{T_n}$ and $V_0^{T_n}$ for which the above random variables are well defined (i.e., they do not evaluate to $\infty-\infty$)  and finite.

Without loss of generality assume that $X_0^{T_n}(1,1)$ was transmitted. Let $P_{e,0}^{(z)}$ denote the probability of the error event that the second decoder  does not find any $V_0^{T_n}$ that satisfies (\ref{EQ:Dec_z_LN}). Due to Lemma \ref{Le:Conv1}, $\E_{\mathcal{C}}[P_{e,0}^{(z)}]$ can be made arbitrarily small, where $\E_{\mathcal{C}}$ denotes expectation with respect to random code book generation.  Let $\mathsf{E}_{e,j}^{(z)}$ denote the    error event that for some $j\neq 1$, $V_0^{T_n}(j)$ satisfies $(\ref{EQ:Dec_z_LN})$, and let $P_{e,j}^{(z)}$ denote the corresponding error probability. Then we have for $j \neq 1$
\begin{align*}
\E_{\mathcal{C}}[P_{e,j}^{(z)}]&=\int_{V_0^{T_n},Z_0^{T_n}} \mathbf{1}\{\mathsf{E}_{e,j}^{(z)}\} d(P^{V_0^{T_n}}\times P^{Z_0^{T_n}})\nn
&\le\exp(-T_n(C_z-\gamma_z))\int_{V_0^{T_n},Z_0^{T_n}}\mathbf{1}\{\mathsf{E}_{e,j}^{(z)}\}d{P}^{V_0^{T_n},Z_0^{T_n}}\nn
&\le\exp(-T_n(C_z-\gamma_z)).
\end{align*}
By the union bound
\begin{align}
\E_{\mathcal{C}}[P_{e}^{(z)}]&\le\E_{\mathcal{C}}[P_{e,0}^{(z)}]+\sum_{j=2}^{L_z}\E_{\mathcal{C}}[P_{e,j}^{(z)}]\nn
&\le\E_{\mathcal{C}}[P_{e,0}^{(z)}]+\exp(-T_n(C_z-R_z-\gamma_z)).
\end{align}
Thus $\E_{\mathcal{C}}[P_{e}^{(z)}]$ can be made arbitrarily small.

Similar to the second decoder, the average probability $\E_{\mathcal{C}}[P_{e,0}^{(y)}]$ that the first receiver cannot find any $(i,j)$ that satisfy both (\ref{EQ:Dec_y11}) and (\ref{EQ:Dec_y21}) can be made small due to Lemma~\ref{Le:Conv1}.
Let $\mathsf{E}_{e,(i,j)}^{(y)}$ denote the error event that for some $(i,j)\neq (1,1)$, $(i,j)$ satisfies both (\ref{EQ:Dec_y11}) and (\ref{EQ:Dec_y21}).
First consider $\mathsf{E}_{e,(i,j)}^{(y)}$ for $j\neq 1$. For this case $X_0^{T_n}(i,j)$ and $Y_0^{T_n}$ are independent, and  for $j\neq 1$, the corresponding error probability $P_{e,(i,j)}^{(y)}$ is upper bounded by the probability that $(i,j)$ satisfies (\ref{EQ:Dec_y11}).
\begin{align*}
\E_{\mathcal{C}}[P_{e,(i,j)}^{(y)}]&\le \int_{X_0^{T_n},Y_0^{T_n}} \mathbf{1}\{\mathsf{E}_{e,(i,j)}^{(y)}\}d(P^{X_0^{T_n}}\times P^{Y_0^{T_n}})\nn
&\le\exp(-T_n(C_y+\tilde{C}_y-\gamma_y))\int_{X_0^{T_n},Y_0^{T_n}}\mathbf{1}\{\mathsf{E}_{e,(i,j)}^{(y)}\}d{P}^{X_0^{T_n},Y_0^{T_n}}\nn
&\le\exp(-T_n(C_y+\tilde{C}_y-\gamma_y)).
\end{align*}

When $j=1$, and $i \neq 1$, $ X_0^{T_n}(i,1)\rightleftarrows V_0^{T_n}(1)\rightleftarrows Y_0^{T_n}$ is a Markov chain. The average probability that $V_0^{T_n}(1)$ and $X_0^{T_n}(i,1)$ for $i\neq 1$ satisfies (\ref{EQ:Dec_y21}) is
\begin{align*}
\int_{V_0^{T_n},X_0^{T_n},Y_0^{T_n}}\mathbf{1}\{\mathsf{E}_{e,(i,1)}^{(y)}\}d{Q}^{V_0^{T_n},X_0^{T_n},Y_0^{T_n}},
\end{align*}
where   ${Q}^{V_0^{T_n},X_0^{T_n},Y_0^{T_n}}$ is defined in (\ref{EQ:Measure_Q}). Thus for $i\neq 1$, we can upper bound $\E_{\mathcal{C}}[P_{e,(i,1)}^{(y)}]$  as
\begin{align*}
\E_{\mathcal{C}}[P_{e,(i,1)}^{(y)}] \le &\int_{V_0^{T_n},X_0^{T_n},Y_0^{T_n}}\mathbf{1}\{\mathsf{E}_{e,(i,1)}^{(y)}\}d{Q}^{V_0^{T_n},X_0^{T_n},Y_0^{T_n}}\nn
&\le\exp(-T_n(C_y-\gamma_y))\int_{V_0^{T_n},X_0^{T_n},Y_0^{T_n}}\mathbf{1}\{\mathsf{E}_{e,(i,1)}^{(y)}\}d{P}^{V_0^{T_n},X_0^{T_n},Y_0^{T_n}}\nn
&\le\exp(-T_n(C_y-\gamma_y)).
\end{align*}
The average probability  of error can be upper bounded using the union bound as
\begin{align}
\E_{\mathcal{C}}[P_{e}^{(y)}]&\le\E_{\mathcal{C}}[P_{e,0}^{(y)}]+\sum_{(i,j)\neq (1,1)}^{L_y,L_z}\E_{\mathcal{C}}[P_{e,(i,j)}^{(y)}]\nn
&=\E_{\mathcal{C}}[P_{e,0}^{(y)}]+\sum_{i=2}^{L_y}\E_{\mathcal{C}}[P_{e,(i,1)}^{(y)}]+\sum_{i=1,j=2}^{L_y,L_z}\E_{\mathcal{C}}[P_{e,(i,j)}^{(y)}]\nn
&=\E_{\mathcal{C}}[P_{e,0}^{(y)}]+(L_y-1)\E_{\mathcal{C}}[P_{e,(2,1)}^{(y)}]+L_y(L_z-1)\E_{\mathcal{C}}[P_{e,(1,2)}^{(y)}]\nn
&\le\E_{\mathcal{C}}[P_{e,0}^{(y)}]+\exp(R_yT_n)\exp(-T_n(C_y-\gamma_y))+\exp((R_y+R_z)T_n)\exp(-T_n({C}_y+\tilde{C}_y-\gamma_y))\nn
&=\E_{\mathcal{C}}[P_{e,0}^{(y)}]+\exp(-T_n(C_y-R_y-\gamma_y))+\exp(-T_n({C}_y+\tilde{C}_y-(R_y+R_z)-\gamma_y)),
\end{align}
which can be made arbitrarily small since $R_y= C_y-\delta$ and 
\begin{align*}
R_y+R_z&=\alpha(p\phi_y(1)+(1-p)\phi_y(0)-\phi_y(p))+(1-\alpha)(q\phi_y(1)+(1-q)\phi_y(0)-\phi_y(q)) \nn
&\quad\,+\alpha\phi_z(p)+(1-\alpha)\phi_z(q)-\phi_z(\alpha p+(1-\alpha)q)-2\delta \nn
&\le\alpha(p\phi_y(1)+(1-p)\phi_y(0)-\phi_y(p))+(1-\alpha)(q\phi_y(1)+(1-q)\phi_y(0)-\phi_y(q))-2\delta \nn
&\quad\,+\alpha\phi_y(p)+(1-\alpha)\phi_y(q)-\phi_y(\alpha p+(1-\alpha)q)\nn
&={C}_y+\tilde{C}_y-2\delta,
\end{align*} 
where we have used the more capable  property of the channel:
 $$\alpha\phi_z(p)+(1-\alpha)\phi_z(q)-\phi_z(\alpha p+(1-\alpha)q)\le\alpha\phi_y(p)+(1-\alpha)\phi_y(q)-\phi_y(\alpha p+(1-\alpha)q).$$

Hence by Markov's inequality, for a given $\epsilon>0$ there exists $N$ and $\bar{\tau}$ such that for all $n\ge N$, and $\tau \le \bar{\tau}$,   a codebook  with $T=n\tau$ satisfying (\ref{EQ:BC_achv}) can be found.

\subsection{Converse}
Suppose that $(R_y,R_z)$ is achievable. Then there exists a code such that (\ref{EQ:BC_achv}) holds.  For $(u,U)\in\{(y,Y),(z,Z)\}$, let $\tilde{R}_u=\frac{\log(L_u)}{T}$. Then 
\begin{eqnarray*}
\tilde{R}_uT=\log(L_u)=H(M_u)&=&\E[H(M_u|U_0^T)]+I(M_u;U_0^T)\\
&\stackrel{(a)}{\le}& H\left(M_u|\mathscr{D}_u^T(U_0^T)\right)+I(M_u;U_0^T)\\
&\stackrel{(b)}{\le}& H(P_e^{(u)})+P_e^{(u)}\log(L_u)+I(M_u;U_0^T).
\end{eqnarray*}
Here $P_e^{(y)}$ and $P_e^{(z)}$ are the average probability of error at the first and second receiver respectively.  Since $M_u\rightleftarrows U_0^T \rightleftarrows \mathscr{D}_u^T(U_0^T) $ is a Markov chain, $I(M_u;U_0^T)\ge I(M_u;\mathscr{D}_u^T(U_0^T))$. Then applying Lemma~\ref{Le:Wyner} gives (a), and 
(b) is an application of Fano's inequality. 
Hence
\begin{align}
 \tilde{R}_u &\leq \frac{1}{T(1-P_e^{(u)})}\left(I(M_u;U_0^T)+H(P_e^{(u)})\right)\nonumber\nn
 &\leq \frac{1}{T(1-\epsilon)}\left(I(M_u;U_0^T)+H(\epsilon)\right).
 \label{EQ:R_u_conv}
\end{align}
Thus
\begin{align}
R_u \le \frac{\log(L_u)}{T}+\epsilon&=\tilde{R}_u+\epsilon\nn
&\le \frac{1}{T(1-\epsilon)}\left(I(M_u;U_0^T)+H(\epsilon)\right)+\epsilon.
\end{align}
Now consider
\begin{eqnarray}
\frac{1}{T}I(M_y;Y_0^T)&\stackrel{}{\le}&\frac{1}{T}I(M_y;M_zY_0^T) \nn
&\stackrel{(a)}{=}&\frac{1}{T}I(M_y;Y_0^T|M_z) \nn
&\stackrel{(b)}{=}&\frac{1}{T}I(M_yM_z;Y_0^T)-\frac{1}{T}I(M_z;Y_0^T) \nn
&\stackrel{(c)}{\le}& \frac{1}{T}I(X^T_0;Y_0^T)-\frac{1}{T}I(M_z;Y_0^T) \nn
&\stackrel{(d)}{=}&\frac{1}{T}I(X^T_0;Y_0^T|M_z)\label{Eq:UBRy1}\\
&\stackrel{(e)}{=}&\frac{1}{T}\int_0^T\left(\mathbb{E}[\phi_y(X_t)]-\mathbb{E}[\phi_y(\mathbb{E}[X_t|Y_0^t,M_z])] \right) \,dt . \nn
&\stackrel{(f)}{=}&\mathbb{E}[\phi_y(X_S)]-\mathbb{E}[\phi_y(\mathbb{E}[X_S|Y_0^S,M_z])]. \label{Eq:UBRy}
\end{eqnarray}
Here, (a) is due to the independence of $M_y$ and $M_z$,  \\*
(b) due to an application of Kolmogrov's formula, \\*
(c) follows since $M_y,M_z\rightleftarrows X_0^T \rightleftarrows Y_0^T $ forms a Markov chain, \\*
(d) follows since $M_z\rightleftarrows X_0^T \rightleftarrows Y_0^T $ forms a Markov chain,  \\*
(e) is an application of Theorem \ref{Th:Mu_inf}, and \\*
(f)  follows by defining $S$ to be a random variable uniformly distributed on $[0,T]$, and independent of all $\sigma$-fields on $(\Omega,\F)$.\footnote{$S$ can be defined by extending the probability space $(\Omega,\F)$ in (\ref{EQ:Prob_space_bc}) to $(\Omega \times [0,T],\F\otimes\mathfrak{B}([0,T]))$, where $\mathfrak{B}([0,T])$ is the Borel $\sigma$-field on $[0, T]$.}

Similarly,
\begin{eqnarray}
\frac{1}{T}I(M_z;Z_0^T)&\stackrel{(a)}{=}&\frac{1}{T}\int_0^T\mathbb{E}[\phi_z(\mathbb{E}[X_t|Z_t^T,M_z])]- \mathbb{E}[\phi_z(\mathbb{E}[X_t|Z_t^T])] \,dt \nn
&\stackrel{(b)}{\le}&\frac{1}{T}\int_0^T\mathbb{E}[\phi_z(\mathbb{E}[X_t|Z_t^T,M_z])] \,dt-\phi_z\left(\frac{1}{T}\int_0^T\mathbb{E}[X_t]\,dt\right) \label{Eq:UBR2}\nn
&\stackrel{(c)}{\le}&\frac{1}{T}\int_0^T\mathbb{E}[\phi_z(\mathbb{E}[X_t|Y_0^t,M_z])]\,dt-\phi_z\left(\frac{1}{T}\int_0^T\mathbb{E}[X_t]\,dt\right) 
\nn
&\stackrel{(d)}{=}&\mathbb{E}[\phi_z(\mathbb{E}[X_S|Y_0^S,M_z])]-\phi_z(\mathbb{E}[X_S]).\label{Eq:UBRz}
\end{eqnarray}
Here, (a) follows from Theorem 1, \\*
(b) from Jensen's inequality applied to the convex function $\phi_z$, \\*
 (c) is due to Lemma \ref{Le:ln_inq}, and \\*
 (d) holds since  $S$ is the random variable, uniformly distributed on $[0,T]$ and independent of all other variables.

\newadd{Since the capacity region is convex, to show that the rate-pair $(R_y, R_z)$ is contained in the region in the statement of the theorem, we use \aaronadd{a} supporting-hyperplane argument. It suffices to show that for any $ \mu_y,\mu_z\ge 0$,
\begin{align*}
\sup_{R_y,R_z} \mu_y R_y+\mu_z R_z \le \sup_{\substack{0\le\alpha\le 1/2 \\
0\le p,q, \le 1}} \mu_y C_y+\mu_z C_z.
\end{align*}
}
 Note that (\ref{EQ:R_u_conv}), (\ref{Eq:UBRy}), and (\ref{Eq:UBRz}) imply
\begin{align}
\mu_y R_y+\mu_z R_z \leq  \mu_y\mathbb{E}[\phi_y(X_S)]- \mu_z \phi_z(\mathbb{E}[X_S])-\mathbb{E}[K_\mu(\mathbb{E}[X_S|Y_0^S,M_z])]+\varepsilon(\epsilon), \label{EQ:rate_mu}
\end{align}
where
\begin{align}
\label{EQ:K_mu}
K_\mu(x)=\mu_y \phi_y(x)-\mu_z \phi_z(x),
\end{align} 
and $\varepsilon(\epsilon)\to 0$ as $\epsilon\to 0$.
We now use Fenchel-Eggleston-Carath\'{e}odory's  theorem \cite[Lemma 15.4, Chapter 15, p. 310]{Salehi}. Since $K_\mu(x)$ is a continuous function, there \newadd{exist $0\le\alpha\le 1/2$}, $0\le p,q \le1$ such that
\begin{align}
\mathbb{E}[K_\mu(\mathbb{E}[X_S|Y_0^S,M_z])]=&\alpha K_\mu(p)+(1-\alpha) K_\mu(q), \\
\mathbb{E}[X_S]=\E[\mathbb{E}[X_S|Y_0^S,M_z]]=&\alpha p+(1-\alpha)q.
\end{align}
Due to the convexity of $\phi_y(x)$ and $0\le X_S \le 1$ with $\mathbb{E}[X_S]=\alpha p+(1-\alpha)q$,
\begin{align}
\mathbb{E}[\phi_y(X_S)] &\le \mathbb{E}[X_S]\phi_y(1)+(1-\mathbb{E}[X_S])\phi_y(0)\nn
&=(\alpha p+(1-\alpha)q)\phi_y(1)+(\alpha(1-p)+(1-\alpha)(1-q))\phi_y(0). \label{EQ:UB_PHIy}
\end{align}
Equations (\ref{EQ:rate_mu})-(\ref{EQ:UB_PHIy}) give
\begin{align}
\mu_y R_y+\mu_z R_z  \leq &  \mu_y((\alpha p+(1-\alpha)q)\phi_y(1)+(\alpha(1-p)+(1-\alpha)(1-q))\phi_y(0))-\mu_z\phi_z(\alpha p+(1-\alpha)q)\nn
&-\alpha (\mu_y\phi_y(p)-\mu_z\phi_z(p))-(1-\alpha) (\mu_y\phi_y(q)-\mu_z\phi_z(q))+\varepsilon(\epsilon)\nn
=&\mu_y\left[(\alpha p+(1-\alpha)q)\phi_y(1)+(\alpha(1-p)+(1-\alpha)(1-q))\phi_y(0)-\alpha\phi_y(p)-(1-\alpha)\phi_y(q) \right] \nn
&+\mu_z \left[\alpha\phi_z(p)+(1-\alpha)\phi_z(q)-\phi_z(\alpha p+(1-\alpha)q) \right]+\varepsilon(\epsilon)\nn
= & \mu_y C_y +\mu_z C_z+\varepsilon(\epsilon)\nn
\le & \newadd{\sup_{\substack{0\le\alpha\le 1/2 \\
0\le p,q, \le 1}} \mu_y C_y+\mu_z C_z + \varepsilon(\epsilon).}
\end{align}
\newadd{Since $\epsilon$ is arbitrary, taking $\epsilon\to 0$  we get the converse part of the theorem.}

\section{More Capable  Poisson Wiretap Channel}
\subsection{Encoding and Decoding }
Here we will consider the first receiver to be the legitimate user and the second receiver to be an eavesdropper. The transmitter (equipped with a stochastic encoder $\mathscr{E}^T_x$) wishes to communicate a message $M$, which is uniformly distributed on $\mathcal{M}=\{1,\dots, L \}$, to the legitimate user (equipped with decoder $\mathscr{D}^T_y$). To transmit message $M=m$, the encoder chooses an input waveform $X_0^T \in \mathcal{X}_0^T$ .  Upon observing $Y_0^T$, the legitimate decoder chooses a symbol $\hat{M}\in\mathcal{M}$. We will call such an arrangement an \emph{$(L,T)$ code}. 
The average probability of error at the legitimate receiver is
\begin{align}
\text{P}_e=\frac{1}{L}\sum_{m=1}^{L} P(\mathscr{D}_y^T(Y_0^T)\neq m | M=m). 
\end{align} 
The metric to measure the secrecy will be $\dfrac{1}{T}I(M;Z_0^T)$.
\begin{Def}
A secrecy rate $R_s$ is said to be \emph{achievable}  for the   Poisson wiretap channel if for all $\epsilon>0$ and for all sufficiently large $T$, there exists an $(L,T)$ code such that 
\begin{align}
\frac{\log(L)}{T}&\ge R_s-\epsilon \nn
\text{P}_e&\le\epsilon \nn
\frac{1}{T}I(M;Z_0^T)&\le\epsilon.
\label{EQ:wiretap_ach} 
\end{align}
\end{Def}
The secrecy capacity is defined to be the supremum of achievable secrecy rate. 
\begin{Theorem}
The secrecy capacity of the more capable  Poisson wiretap channel is 
\begin{align}
C_s=\max_{0\le\alpha\le 1}\alpha\Phi(1)+(1-\alpha)\Phi(0)-\Phi(\alpha),
\end{align}
where we recall $\Phi(x)=\phi_y(x)-\phi_z(x)$ and $\Phi(x)$ is a convex function. 
\end{Theorem}
Note that  this capacity expression is same as  that of the capacity of the degraded Poisson wiretap channel in \cite{Laourine}.  Since the  achievability argument is identical to that for  the degraded Poisson wiretap channel in \cite[Section III]{Laourine}, we shall only prove the converse here.
\subsection{Converse}
Suppose $R_s$ is achievable. Then there exists an $(L,T)$ code satisfying (\ref{EQ:wiretap_ach}).
Let $R=\frac{\log(L)}{T}$, then
\begin{align*}
RT=\log(L)=H(M)=&\E\left[H(M|Y_0^T)\right]+I(M;Y_0^T)\\
\stackrel{(a)}{\le}& H\left(M|\mathscr{D}_y^T(Y_0^T)\right)+I(M;Y_0^T)\\
\stackrel{(b)}{\le}& H(\text{P}_e)+\text{P}_e\log(L)+I(M;Y_0^T).
\end{align*}
Since $M\rightleftarrows Y_0^T \rightleftarrows \mathscr{D}_y^T(Y_0^T) $ is a Markov chain, $I(M;Y_0^T)\ge I(M;\mathscr{D}_y^T(Y_0^T))$. Then applying Lemma  \ref{Le:Wyner} gives (a), and 
(b) is an application of Fano's inequality. 
This gives 
\begin{align*}
R&\le\frac{1}{T(1-P_e)}(I(M;Y_0^T)+H(\text{P}_e)) \\
&=\frac{1}{T(1-P_e)}(I(M;Y_0^T)-I(M;Z_0^T)+H(\text{P}_e)+I(M;Z_0^T))\\
&\le\frac{1}{T(1-\epsilon)}((I(M;Y_0^T)-I(M;Z_0^T)+H(\epsilon))+\frac{\epsilon}{1-\epsilon}.
\end{align*}
Now consider
\begin{align*}
\frac{1}{T}\left(I(M;Y_0^T)-I(M;Z_0^T)\right)
\stackrel{(a)}{=}&\frac{1}{T}\int_0^T\mathbb{E}[\phi_y(\mathbb{E}[X_t|Y_0^t,M])]- \mathbb{E}[\phi_y(\mathbb{E}[X_t|Y_0^t])] \,dt\\
&-\frac{1}{T}\int_0^T\mathbb{E}[\phi_z(\mathbb{E}[X_t|Z_t^T,M])]- \mathbb{E}[\phi_z(\mathbb{E}[X_t|Z_t^T])]\,dt \nn
=&\frac{1}{T}\int_0^T\mathbb{E}[\phi_y(\mathbb{E}[X_t|Y_0^t,M])]-\mathbb{E}[\phi_z(\mathbb{E}[X_t|Z_t^T,M])]\,dt\\
&-\frac{1}{T}\int_0^T\mathbb{E}[\phi_y(\mathbb{E}[X_t|Y_0^t])]-\mathbb{E}[\phi_z(\mathbb{E}[X_t|Z_t^T])]\,dt\\
\stackrel{(b)}{=}&\frac{1}{T}\int_0^T\mathbb{E}[\phi_y(\mathbb{E}[X_t|Y_0^t,Z_t^T,M])]-\mathbb{E}[\phi_z(\mathbb{E}[X_t|Y^t_0,Z_t^T,M])]\,dt\\
&-\frac{1}{T}\int_0^T\mathbb{E}[\phi_y(\mathbb{E}[X_t|Y_0^t,Z_t^T])]-\mathbb{E}[\phi_z(\mathbb{E}[X_t|Y_0^t,Z_t^T])]\,dt\\
\stackrel{}{=}&\frac{1}{T}\int_0^T\mathbb{E}[\Phi(\mathbb{E}[X_t|Y_0^t,Z_t^T,M])]-\mathbb{E}[\Phi(\mathbb{E}[X_t|Y^t_0,Z_t^T])]\,dt\\
\stackrel{(c)}{\le}& \frac{1}{T}\int_0^T\mathbb{E}[\Phi(X_t)]-\Phi(\mathbb{E}[X_t])\,dt\\
\stackrel{(d)}{=}&  \max_{0\le\alpha\le 1}\alpha\Phi(1)+(1-\alpha)\Phi(0)-\Phi(\alpha)\\
\stackrel{}{=}&\,\, C_s.
\end{align*}
Here, for (a) we have used Theorem \ref{Th:Mu_inf},\\*
for (b) we have used Theorem \ref{Thm:Integral_Identity},\\*
for (c) we have applied Jensen's inequality to both terms in the integral, and \\*
(d) follows  from fixing the mean of the input distribution to $\alpha$ and  maximizing over all such distributions and then maximizing over $\alpha$. Due to the convexity of $\Phi(x)$,  the maximizing distribution puts mass on the extreme points $\{0,1\}$, that is, mass $1-\alpha$ on $0$ and mass $\alpha$ on $1$.

Hence we get,
\begin{align*}
R_s &\le \frac{\log(L)}{T}+\epsilon \nn
&\le\frac{C_s}{1-\epsilon}+\frac{H(\epsilon)}{T(1-\epsilon)}+\frac{\epsilon}{1-\epsilon}.
\end{align*}
Since $\epsilon$ is arbitrary, taking $\epsilon\to 0$ we get the converse part of the theorem.

\section{General Poisson Broadcast Channel with Degraded Message Sets}
In this setting the transmitter has a common message $M_o\in\mathcal{M}_0=\{1,\dots,L_0\}$ for both of the users and a private message $M_y\in\mathcal{M}_y=\{1,\dots,L_y\}$ for the first user. Messages $M_0$ and $M_y$ are assumed to be independent and uniformly distributed on their respective support. The transmitter uses an encoder $\mathscr{E}_x^T$ which maps these messages into an input $X_0^T$
\begin{align*}
\mathscr{E}_x^T:{\M_0\times\M_y}\to\mathcal{X}_0^T.
\end{align*}
Upon observing $Y_0^T$, the first receiver estimates both common and private messages using decoder $\mathscr{D}_y^T$
\begin{align*}
\mathscr{D}_y^T:\mathcal{N}_0^T\to{\M_0\times\M_y}.
\end{align*}
Similarly the second receiver employs  $\mathscr{D}_z^T$ to decode the common message
\begin{align*}
\mathscr{D}_z^T:\mathcal{N}_0^T\to{\M_0}.
\end{align*}
We will call the above setup an \emph{$(L_0,L_y,T)$ code}. The average probability of error of this code is 
\begin{align*}
\text{P}_e=\frac{1}{L_0L_y}\sum_{m_0=1,m_y=1}^{L_0,L_y}P\left\{\{\mathscr{D}^T_{y}(Y_0^T)\neq (m_0,m_y)\} \bigcup \{\mathscr{D}^T_{z}(Z_0^T)\neq m_0\} \middle| M_0=m_0,M_y=m_y \right\}.
\end{align*}
The rate pair $(R_0,R_y)$ is said to be \emph{achievable} if for any $\epsilon >0$ and for any sufficiently large $T$, there exists an $(L_0,L_y,T)$ code such that 
\begin{align}
\frac{\log(L_0)}{T}&\ge R_0-\epsilon \nn
\frac{\log(L_y)}{T}&\ge R_y-\epsilon \nn
\text{P}_e&\le\epsilon. 
\end{align}
The capacity region is the closure of the achievable rate pairs.
Let $P_{e,0}^{(y)}$, $P_{e,y}^{(y)}$ denote the average probability of \newadd{error in} decoding messages $M_0$ and $M_y$, respectively, at the first receiver and similarly let $P_{e,0}^{(z)}$ denote the average probability of error at the second receiver. Then for a given code
\begin{align}
\max(P_{e,0}^{(y)},P_{e,y}^{(y)},P_{e,0}^{(z)}) \le \text{P}_e \le P_{e,0}^{(y)}+P_{e,y}^{(y)}+P_{e,0}^{(z)}.
\end{align}
 
\begin{Theorem}
\label{Thm:Cap_deg}
The capacity region of the general Poisson broadcast channel with degraded message sets is given by the union over all $0\le\alpha_i,p_i\le 1$, $i=1,2,3$ with $\sum\limits_{i=1}^3\alpha_i=1$ of rate pairs satisfying:
\begin{align*}
R_0&\le C_z\nn
R_0+R_y&\le \hat{C}_y+\tilde{C}_y \nn
R_0+R_y&\le C_z+\hat{C}_y,
\end{align*}
where
\begin{align*}
C_z&=\sum_{i=1}^3\alpha_i \phi_z(p_i)-\phi_z\left(\sum_{i=1}^3\alpha_i p_i\right) \nn
\hat{C}_y&=\sum_{i=1}^3 \alpha_i \left(p_i\phi_y(1)+(1-p_i)\phi_y(0)-\phi_y(p_i)\right) \nn
\tilde{C}_y&=\sum_{i=1}^3\alpha_i \phi_y(p_i)-\phi_y\left(\sum_{i=1}^3\alpha_i p_i\right).
\end{align*}
\end{Theorem}

\subsection{Achievability}
\newadd{We will show the achievability of the \aaronadd{formally} larger region:}
\begin{align}
R_y&\le \hat{C}_y\nn
R_0& \le C_z\nn
R_0+R_y&\le \hat{C}_y+\tilde{C}_y.
\label{EQ:deg_ach_reg}
\end{align}
\aaronadd{The above region turns out to equal the region in the statement of the theorem, which will follow from the converse proven later. To see that the region in (\ref{EQ:deg_ach_reg}) indeed contains the one given in the theorem, it suffices to show that} the rate pair $\bar{R}_0=C_z>0$ and $\bar{R}_y=\min((\hat{C}_y+\tilde{C}_y-C_z),\hat{C}_y)>0$ is in (\ref{EQ:deg_ach_reg}). This follows since
$\bar{R}_0$ and $\bar{R}_y$ satisfy
\begin{align*}
\bar{R}_y\le\hat{C}_y\nn
\bar{R}_0 = C_z \nn
\bar{R}_0+\bar{R}_y \le \hat{C}_y+\tilde{C}_y.
\end{align*}

We use superposition coding and a similar argument as that used in the achievability proof for the more capable Poisson broadcast channel with independent message sets. We divide the interval $[0, T_n]$ into $n$ intervals each of equal length $\tau=T_n/n$. Here we take $V_0^{T_n}$ to be a ternary stochastic process.  The process will be constant on each of these sub-interval with value given by 
\begin{align}
V_t=\bar{V}_i \mbox{ for } (i-1)\tau \le t < i\tau, \quad i=1,2,\dots,n
\label{EQ:deg_V1}
\end{align}
where $\bar{V}_i$ are independent and identically distributed  random variables with 
\begin{align}
P(\bar{V}_i=j)=\alpha_j, j\in\{1,2,3\}.
\label{EQ:deg_V2}
\end{align}
We  construct the input processes, $X_0^T$, as binary and piecewise constant with
\begin{equation}
X_t=\bar{X}_i \mbox{ for } (i-1)\tau \le t < i\tau, \quad i=1,2,\dots,n,
\label{EQ:deg_X1}
\end{equation}
and
\begin{align}
P(\bar{X}_i=1|\bar{V}_i=j)=1-P(\bar{X}_i=0|\bar{V}_i=j)=p_j, j\in\{1,2,3\}.
\label{EQ:deg_X2}
\end{align}
Lemma \ref{Le:Conv} gives that for all $\epsilon>0$ there exists $\bar{\tau}$ and $N$ such that if $n\ge N$ and $\tau \le \bar{\tau}$ then 
\begin{align}
P\left(\left|\frac{1}{T_n}\ii(V_0^{T_n};Z_0^{T_n})-C_z\right|>\epsilon\right)\le\epsilon\nn
P\left(\left|\frac{1}{T_n}\ii(X_0^{T_n};Y_0^{T_n})-(C_y+\tilde{C}_y\right)|>\epsilon\right)\le\epsilon\nn
P\left(\left|\frac{1}{T_n}\ii(X_0^{T_n};Y_0^{T_n}|V_0^{T_n})-C_y\right|>\epsilon\right)\le\epsilon.
\end{align}
\subsubsection*{Encoding and Decoding Operation}
Let $({R}_0,{R}_y)$ be strictly positive, satisfying (\ref{EQ:deg_ach_reg}), and let $\tilde{R}_u={R}_u-\delta$, $u\in\{0,y\}$ for some $\delta>0$.
We generate $L_0=\exp(T_n\tilde{R}_0)$ many $V_0^{T_n}$ waveforms (indexed by $j=1,\dots, L_0 $) independently according to (\ref{EQ:deg_V1}) and (\ref{EQ:deg_V2}). For  each $V_0^{T_n}(j)$,  we generate $L_y=\exp(T_n\tilde{R}_y)$ many independent $X_0^{T_n}$ waveforms (indexed by $i=1,\dots,L_y$) according to (\ref{EQ:deg_X1}) and (\ref{EQ:deg_X2}). To transmit messages $(M_0,M_y)$, the encoder sends $X_0^{T_n}(M_0,M_y)$ over the channel.  

Both of the receivers consider only those inputs for which the mutual information densities (in Definition \ref{Def:MID}) evaluate to a finite value (computed using Theorem \ref{Th:Mu_inf}) for given received point process. 
The first receiver seeks unique $i$ and $j$ that satisfy both
\begin{align}
\frac{1}{{T_n}}\ii(X_0^{T_n}(i,j);Y_0^{T_n})\ge {C}_y+\tilde{C}_y-\gamma_y
\label{EQ:Dec_y1}
\end{align}
and
\begin{align}
\frac{1}{{T_n}}\ii(X_0^{T_n}(i,j);Y_0^{T_n}|V_0^{T_n}(j))\ge C_y-\gamma_y.
\label{EQ:Dec_y2}
\end{align}
The second decoder finds the unique $j$   such that 
\begin{align}
\frac{1}{{T_n}}\ii(V_0^{T_n}(j);Z_0^{T_n})\ge C_z-\gamma_z
\label{EQ:Dec_z}
\end{align}
for some $\gamma_z>0$.
Without loss of generality assume that $X_0^T(1,1)$ was transmitted over the channel.
Using a similar argument as that for the error analysis in the achievability proof of the more capable channel with independent messages we get the following.
Since
\begin{align}
\tilde{R}_0+\tilde{R}_y&= C_y+\tilde{C}_y-2\delta\nn
\tilde{R}_y&= C_y-\delta,
\end{align}
the expectation (over random codebook generation) of the average probability of error at the first receiver can be made arbitrarily small. 
 Similarly, as $\tilde{R}_0=C_z-\delta$, the expectation of the average probability of error at the second receiver can be made arbitrarily low. 
Hence there exists a sequence of codebooks which achieve the rates in (\ref{EQ:deg_ach_reg}) with arbitrarily low probability of error. 
\subsection{Converse}
For a given sequence of $(L_0,L_y,T)$ codes, using Lemma \ref{Le:Wyner} and Fano's inequality,  we get
\begin{align*}
R_0 &\leq  \frac{1}{T(1-\epsilon)}\left(I(M_0;Z_0^T)+H(\epsilon)\right)+\epsilon \nn
R_y &\leq  \frac{1}{T(1-\epsilon)}\left(I(M_y;Y_0^T)+H(\epsilon)\right)+\epsilon \nn
R_0+R_y &\leq \frac{1}{T(1-\epsilon)}\left(I(M_0,M_y;Y_0^T)+H(\epsilon)\right)+2\epsilon,
\end{align*}
where we have used the fact that the first user needs to decode both $M_0$ and $M_y$, whereas second receiver requires only $M_0$.
We now upper bound the mutual information expressions in the above inequalities.
\begin{align}
\frac{1}{T}I(M_0;Z_0^T)  
&\stackrel{(a)}{=}\frac{1}{T}\int_0^T\mathbb{E}[\phi_z(\mathbb{E}[X_t|Z_t^T,M_0])]- \mathbb{E}[\phi_z(\mathbb{E}[X_t|Z_t^T])] \,dt \nn
&\stackrel{(b)}{\le}\frac{1}{T}\int_0^T\mathbb{E}[\phi_z(\mathbb{E}[X_t|Z_t^T,M_0])] \,dt-\phi_z\left(\frac{1}{T}\int_0^T\mathbb{E}[X_t]\,dt\right) \nn
&\stackrel{(c)}{\le}\frac{1}{T}\int_0^T \mathbb{E}[\phi_z(\mathbb{E}[X_t|Y_0^t,Z_t^T,M_0])] \,dt-\phi_z\left(\frac{1}{T}\int_0^T\mathbb{E}[X_t]\,dt\right) \nn
&\stackrel{(d)}{=}\mathbb{E}[\phi_z(\mathbb{E}[X_S|Z_S^T,Y_0^S,M_0])]-\phi_z(\mathbb{E}[X_S]).
\label{EQ:R_0_deg}
\end{align}
In (a), we have used Theorem \ref{Th:Mu_inf},\\*
in (b) and (c), we have applied Jensen's inequality to the second and first terms in the integrand, respectively, and\\*
in (d), we have defined $S$ to be a random variable, uniformly distributed on $[0, T]$ and independent of all other random variables and processes.
Now consider $\frac{1}{T}I(M_0,M_y;Y_0^T)$.
\begin{align}
\frac{1}{T}I(M_0,M_y;Y_0^T) 
&\stackrel{(a)}{\le}\frac{1}{T}I(X_0^T;Y_0^T) \nn
&\stackrel{}{=}\frac{1}{T}\int_0^T\mathbb{E}[\phi_y(X_t)]- \mathbb{E}[\phi_y(\mathbb{E}[X_t|Y_0^t])] \,dt \nn
&\stackrel{(b)}{\le}\frac{1}{T}\int_0^T \mathbb{E}[\phi_y(X_t)]\,dt- \phi_y\left(\frac{1}{T}\int_0^T\mathbb{E}[X_t]\,dt\right) \nn
&\stackrel{(c)}{=}\mathbb{E}[\phi_y(X_S)]-\phi_y\left(\mathbb{E}[X_S]\right).
\label{EQ:sum_rate_deg1}
\end{align}
Here (a) is due to the Markov chain $(M_0,M_y) \rightleftarrows X_0^T \rightleftarrows Y_0^T$, \\*
(b) is due Jensen's inequality, and \\*
(c) follows because $S$ is a uniformly distributed on $[0,T]$.

Similar to (\ref{Eq:UBRy1}), we can show 
\begin{align*}
R_y \le \frac{1}{T(1-\epsilon)}(I(X_0^T;Y_0^T|M_0)+H(\epsilon))+\epsilon.
\end{align*} 
Now consider
\begin{align}
\frac{1}{T}I(X_0^T;Y_0^T|M_0)+\frac{1}{T}I(M_0;Z_0^T) 
\stackrel{(a)}{=}&\frac{1}{T}\int_0^T\mathbb{E}[\phi_y(X_t)]-\mathbb{E}[\phi_y(\mathbb{E}[X_t|Y_0^t,M_0])]\,dt  \nn
&+\frac{1}{T}\int_0^T \mathbb{E}[\phi_z(\mathbb{E}[X_t|Z_t^T,M_0])]- \mathbb{E}[\phi_z(\mathbb{E}[X_t|Z_t^T])] \,dt \nn
\stackrel{}{=}&\frac{1}{T}\int_0^T \mathbb{E}[\phi_y(X_t)]-\mathbb{E}[\phi_z(\mathbb{E}[X_t|Z_t^T])] \,dt  \nn
&+\frac{1}{T}\int_0^T \mathbb{E}[\phi_z(\mathbb{E}[X_t|Z_t^T,M_0])]- \mathbb{E}[\phi_y(\mathbb{E}[X_t|Y_0^t,M_0])]  \,dt  \nn
\stackrel{(b)}{\le}&\frac{1}{T}\int_0^T\mathbb{E}[\phi_y(X_t)] \,dt-\phi_z\left(\frac{1}{T}\int_0^T\mathbb{E}[X_t]\,dt\right)  \nn
&+\frac{1}{T}\int_0^T\mathbb{E}[\phi_z(\mathbb{E}[X_t|Z_t^T,M_0])]- \mathbb{E}[\phi_y(\mathbb{E}[X_t|Y_0^t,M_0])]  \,dt  \nn
\stackrel{(c)}{=}&\frac{1}{T}\int_0^T\mathbb{E}[\phi_y(X_t)] \,dt-\phi_z\left(\frac{1}{T}\int_0^T\mathbb{E}[X_t]\,dt\right)  \nn
&+\frac{1}{T}\int_0^T\mathbb{E}[\phi_z(\mathbb{E}[X_t|Y_0^t,Z_t^T,M_0])]- \mathbb{E}[\phi_y(\mathbb{E}[X_t|Y_0^t,Z_t^T,M_0])]  \,dt  \nn
\stackrel{(d)}{=}&\:\mathbb{E}[\phi_y(X_S)]-\phi_z\left(\mathbb{E}[X_S]\right)  \nn*
&+\mathbb{E}[\phi_z(\mathbb{E}[X_S|Y_0^S,Z_S^T,M_0])]- \mathbb{E}[\phi_y(\mathbb{E}[X_S|Y_0^S,Z_S^T,M_0])].
\label{EQ:sum_rate_deg2}
\end{align}
Here, (a) is due to Theorem \ref{Th:Mu_inf},\\
(b) is due to Jensen's inequality,\\ 
(c) is due to Theorem \ref{Thm:Integral_Identity}, and\\ 
(d) follows because $S$ is uniformly distributed on $[0,T]$ and independent of all other random variables.

Now we use Fenchel-Eggleston-Carath\'{e}odory's  theorem \cite[Lemma 15.4, Chapter 15, p. 310]{Salehi}. Since $\phi_y(x)$ and $\phi_z(x)$ are continuous functions,  there exist  $0 \le p_1, p_2, p_3 \le 1$ and $0 \le \alpha_1, \alpha_2, \alpha_3 \le 1$ with $\sum_{i=1}^3 \alpha_i=1$ such that
\begin{align}
\mathbb{E}[\phi_y(\mathbb{E}[X_S|Z_S^T,Y_0^S,M_0])]=&\sum_{i=1}^3\alpha_i \phi_y(p_i) \nn
\mathbb{E}[\phi_z(\mathbb{E}[X_S|Z_S^T,Y_0^S,M_0])]=&\sum_{i=1}^3\alpha_i \phi_z(p_i)\nn
\mathbb{E}[X_S]=\mathbb{E}[\mathbb{E}[X_S|Z_S^T,Y_0^S,M_0]]=&\sum_{i=1}^3\alpha_i p_i.
\end{align}
Due to the convexity of $\phi_u$,
\begin{align}
\E[\phi_u(X_S)]&\le \E[X_S]\phi_u(1)+(1-\E[X_S])\phi_u(0)\nn
&=\sum_{i=1}^3\alpha_i p_i\phi_u(1)+\left(1-\sum_{i=1}^3\alpha_i p_i\right)\phi_u(0)\nn
&=\sum_{i=1}^3\alpha_i \left(p_i\phi_u(1)+(1-p_i)\phi_u(0)\right).
\end{align}
Substituting we get the following.  From (\ref{EQ:R_0_deg})
\begin{align*}
R_0&\le\mathbb{E}[\phi_z(\mathbb{E}[X_S|Z_S^T,Y_0^S,M_0])]-\phi_z(\mathbb{E}[X_S])+\varepsilon(\epsilon) \\
&=\sum_{i=1}^3\alpha_i \phi_z(p_i)-\phi_z\left(\sum_{i=1}^3\alpha_i p_i\right)+\varepsilon(\epsilon)\\
&=C_z+\varepsilon(\epsilon)
\end{align*}
where $\varepsilon(\epsilon)\to 0$ as $\epsilon \to 0$. 
From (\ref{EQ:sum_rate_deg1}) we get
\begin{align*}
R_0+R_y&\le\mathbb{E}[\phi_y(X_S)]-\phi_y\left(\mathbb{E}[X_S]\right)+\varepsilon^\prime(\epsilon)\\
&\le \sum_{i=1}^3\alpha_i \left(p_i\phi_y(1)+(1-p_i)\phi_y(0)\right)-\phi_y\left(\sum_{i=1}^3\alpha_i p_i\right)+\varepsilon^\prime(\epsilon)\nn
&=\hat{C}_y+\tilde{C}_y+\varepsilon^\prime(\epsilon).
\end{align*}
where $\varepsilon^\prime(\epsilon)\to 0$ as $\epsilon \to 0$. Finally (\ref{EQ:sum_rate_deg2}) gives
\begin{align*}
R_0+R_y &\le \mathbb{E}[\phi_y(X_S)]-\phi_z\left(\mathbb{E}[X_S]\right)
+\mathbb{E}[\phi_z(\mathbb{E}[X_S|Y_0^S,Z_S^T,M_0])]- \mathbb{E}[\phi_y(\mathbb{E}[X_S|Y_0^S,Z_S^T,M_0])]+\varepsilon(\epsilon)^{\prime\prime}\\
&\le\sum_{i=1}^3\alpha_i \left(p_i\phi_y(1)+(1-p_i)\phi_y(0)\right)-\phi_z\left(\sum_{i=1}^3\alpha_i p_i\right)\\
&\quad +\sum_{i=1}^3\alpha_i \phi_z(p_i)-\sum_{i=1}^3\alpha_i \phi_y(p_i)+\varepsilon^{\prime\prime}(\epsilon)\\
&=\hat{C}_y+C_z+\varepsilon^{\prime\prime}(\epsilon),
\end{align*}
where $\varepsilon^{\prime\prime}(\epsilon)\to 0$ as $\epsilon \to 0$. As $\epsilon$ is arbitrary, taking $\epsilon \to 0$ completes the converse argument.

\appendix[Proofs of Lemmas]
\begin{IEEEproof}[Proof of Lemma \ref{LE:DSPP}]
Let $[s, t]\in[0,T]$, and $k\in \mathbf{N}$ then
\begin{align}
P(\tilde{\N}_t-\tilde{\N}_s=k|{\tilde{X}}_0^T)&=P(\N_{(T-s)-}-\N_{(T-t)-}=k|X_0^T)\nn
&=\frac{1}{k!}\left(\int_{T-t}^{T-s} X_\tau \,d\tau\right)^k\exp\left(-\int_{T-t}^{T-s} X_\tau \,d\tau\right)\nn
&=\frac{1}{k!}\left(\int_t^s \tilde{X}_\tau \,d\tau\right)^k\exp\left(-\int_t^s \tilde{X}_\tau \,d\tau\right),
\end{align}
where we have used the fact that since $X_0^T$ is c\`adl\`ag, the set $\{t:X_{t-}\neq X_t, \tin\}$ is at most countable \cite[Section 12, Lemma 1, p. 122]{Billingsley}.
Since the new process $\tilde{\N}_0^T$ is obtained by time reversing the process $\N_0^T$, it has the independent increment property. 
\end{IEEEproof}

\begin{IEEEproof}[Proof of Lemma \ref{Le:MC_Int}]
For $0\le s < t \le {T}$
\begin{align}
\E[\hat{\N}_t-\hat{\N}_s|\F_s]&=\E[\hat{\N}_t-\hat{\N}_s|A,\Lambda_0^T,\hat{N}_0^s] \nn
&\stackrel{(a)}{=}\E[\hat{\N}_t-\hat{\N}_s|\Lambda_0^T,\hat{N}_0^s]\nn
&\stackrel{(b)}{=}\int_s^t\hat{\Lambda}_u\,du \nn
&\stackrel{(c)}{=}\E\left[\int_s^t\hat{\Lambda}_u \,du \big|\F_s\right]\label{EQ:MC_INT}.
\end{align}
Here, (a) is due to the fact that if  $A\rightleftarrows \Lambda_0^T \rightleftarrows (\hat{N}_0^{{s}},\hat{N}_s^{{T}})$ is a Markov chain then so is $A\rightleftarrows (\Lambda_0^T,\hat{N}_0^s)\rightleftarrows \hat{N}_s^{{T}}$ \cite[Proposition 6.8, p.111]{Kallenberg}, and then using \cite[Proposition 6.6, p.111]{Kallenberg}, \\*
(b) is due to Definition \ref{Def:DSPP} and the independent increment property of Poisson processes, and \\*
(c) is due to the fact that $\Lambda_0^T$ is measurable with respect to $\F_t$ for all $t\in[0, {T}]$.

Then from (\ref{EQ:MC_INT}) and \cite[Chapter II, Section 2, p. 23-24]{Bremaud} we get that for all non-negative $(\F_t:\tin)$-predictable processes $C_0^{{T}}$
\begin{align}
\E\left[\int_0^{T} C_s \, d\hat{\N}_{s}\right]=\E\left[\int_0^{T} C_s \hat{\Lambda}_s \, ds\right].
\label{EQ:intensity1} 
\end{align}
Also, $\hat{\Lambda}_0^{{T}}$ is $\F_0$-measurable and thus $(\F_t:\tin)$-predictable. Hence the $\left(P,\F_t:\tin\right)$-intensity of $\hat{N}_0^{{T}}$ is $\hat{\Lambda}_0^{{T}}$.

Let ${D}_0^T$ be a non-negative $(\G_t:\tin)$-predictable process. As $\G_t\subseteq\F_t$, it is also $(\F_t:\tin)$-predictable.
Hence 
\begin{align}
\E\left[\int_0^{T} D_s \, d\hat{\N}_{s}\right]=\E\left[\int_0^{T} D_s \hat{\Lambda}_s \, ds\right].
\label{EQ:intensity2}  
\end{align}
Let $\Pi_t=\E[\hat{\Lambda}_t|\G_{t-}]$, $\tin$. Then the process $\Pi_0^T$ is $(\G_t:\tin)$-predictable \cite[Chapter 6, Theorem 43, p. 103]{Dellacherie}.
Hence
\begin{align*}
\E\left[\int_0^{T} D_s \Pi_s \, ds\right]&=\E\left[\int_0^{T} D_s \E[\hat{\Lambda}_s|\G_{s-}] \, ds\right]\nn
&\stackrel{(a)}{=}\E\left[\int_0^{T} \E[D_s \hat{\Lambda}_s|\G_{s-}] \, ds\right]\nn
&=\E\left[\int_0^{T} D_s \hat{\Lambda}_s \, ds\right]\nn
&\stackrel{(b)}{=}\E\left[\int_0^{T} D_s \, d\hat{\N}_{s}\right].
\end{align*}
Here, (a) is due to the fact that $D_s$ is $\G_{s-}$ measurable \cite[Exercise E10, Chapter I, p. 9]{Bremaud}, and\\*
(b) is due to (\ref{EQ:intensity2}).

Hence the $\left(P,\G_t:\tin\right)$-intensity of $\hat{N}_0^{{T}}$ is $\Pi_0^{{T}}$. Since for each $\tin$, $\hat{\N}_{t-}=\hat{\N}_{t}$ $P$-a.s., we can take
\begin{align*}
\Pi_t=\E[\hat{\Lambda}_t|\G_{t-}]=\E[\hat{\Lambda}_t|\G_{t}] \quad P\text{-a.s.}
\end{align*}
\end{IEEEproof}

\begin{IEEEproof}[Proof of Lemma \ref{Le:Abs_cont}]
Using the data processing inequality
\begin{align*}
I(A;\hat{U}_0^T)&=I(A;\U_{t_1}^{t_2})\le I(X_0^T;\U_{t_1}^{t_2})\nn
&\le I(X_0^T;U_0^T)<\infty,
\end{align*}
where the last inequality is due to \cite{Kabanov,Davis,Wyner}.
Hence $P^{A,\hat{U}_0^T} \ll P^A\times P^{\hat{U}_0^T}$.

From (\ref{EQ:Prob_Space}) we get that $P^{{U}_0^T}\ll P_0^{{U}_0^T}$. Let $\mathsf{N}$ be such that $P_0^{\hat{U}_0^T}(\mathsf{N})=0$. Then $P_0^{\hat{U}_0^T}(\mathsf{N})=P_0^{{U}_0^T}((\hat{U}_0^T)^{-1}\mathsf{N})=0$. Hence $P^{{U}_0^T}((\hat{U}_0^T)^{-1}\mathsf{N})=P^{\hat{U}_0^T}(\mathsf{N})=0$. Thus
\begin{align*}
P^{\hat{U}_0^T} \ll P_0^{\hat{U}_0^T}.
\end{align*}
This gives $P^A\times  P^{{\hat{U}_0^T}} \ll P^A\times P_0^{\hat{U}_0^T}$ \cite[Chapter 1, Exercise 19, p. 22]{Kallenberg}.
\end{IEEEproof}

\newadd{
\begin{IEEEproof}[Proof of Lemma \ref{le:intensity_proof}]
Recall that $L_0^T$ can be written as
\begin{align*}
L_t=\exp\left(\int_{0}^t\log(\Psi_s)d\hat{{U}}_s+(1-\Psi_s)\mu_s\,ds\right).
\end{align*}
We note that for $\tin$ $L_t$ satisfies
\begin{align}
L_t=\begin{cases}
L_{t-} & \text{if }\hat{{U}}_t-\hat{{U}}_{t-} =0,\\
\Psi_tL_{t-} & \text{if }\hat{{U}}_t-\hat{{U}}_{t-} =1.
\end{cases}
\label{eq:app1}
\end{align}
Let $C_0^T$ be a non-negative $(\G_t:\tin)$-predictable process. Then
\begin{align*}
\E\left[\int_0^T C_t\,d\hat{{U}}_t\right]&\overset{(a)}{=}\E_{\tilde{P}^{A,\hat{U}_0^T}}\left[L_T\int_0^T  C_t\,d\hat{{U}}_t\right]\\
&\overset{(b)}{=}\E_{\tilde{P}^{A,\hat{U}_0^T}}\left[\int_0^T  L_tC_t\,d\hat{{U}}_t\right]\\
&\overset{(c)}{=}\E_{\tilde{P}^{A,\hat{U}_0^T}}\left[\int_0^T  \Psi_tL_{t-}C_t\,d\hat{{U}}_t\right]\\
&\overset{(d)}{=}\E_{\tilde{P}^{A,\hat{U}_0^T}}\left[\int_0^T  \Psi_tL_{t-}C_t\mu_t\,dt\right]\\
&\overset{(e)}{=}\E_{\tilde{P}^{A,\hat{U}_0^T}}\left[\int_0^T  \Psi_tL_{t}C_t\mu_t\,dt\right]\\
&\overset{(f)}{=}\E_{\tilde{P}^{A,\hat{U}_0^T}}\left[\int_0^T  \hat{\Psi}_tL_{t}C_t\,dt\right]\\
&\overset{(g)}{=}\E_{\tilde{P}^{A,\hat{U}_0^T}}\left[L_T\int_0^T  \hat{\Psi}_tC_t\,dt\right]\\
&\overset{(h)}{=}\E_{}\left[\int_0^T  \hat{\Psi}_tC_t\,dt\right],
\end{align*}
where, (a) follows since $L_T$ is the Radon-Nikodym derivative $\frac{dP^{A,\hat{U}_0^T}}{d\tilde{P}^{A,\hat{U}_0^T}}$,\\*
(b) follows due to~\cite[T19 Theorem, Appendix A2, p. 302]{Bremaud},\\*
(c) follows due to~(\ref{eq:app1}),\\*
(d) follows since the $(\tilde{P}^{A,\hat{U}_0^T},\G_t:\tin)$-intensity of $\hat{U}_0^T$ is $\mu_0^T$, and $L_{t-}$ being a left-continuous adapted process is  $(\G_t:\tin)$-predictable,\\*
(e) follows since the Lebesgue measure of the set $\{t:\tin,L_{t-}\neq L_t\}$ is zero due to~(\ref{eq:app1}),\\*
(f) follows from the definition $\hat{\Psi}_t=\Psi_t\mu_t$,\\*
(g) again follows  due to~\cite[T19 Theorem, Appendix A2, p. 302]{Bremaud},\\*
(h) again follows  since $L_T$ is the Radon-Nikodym derivative $\frac{dP^{A,\hat{U}_0^T}}{d\tilde{P}^{A,\hat{U}_0^T}}$.
\end{IEEEproof}
}

\begin{IEEEproof}[Proof of Lemma \ref{Le:Ch_Rule} ]
Let \begin{equation}
f(t)=\mathbb{E}[\phi_u(\mathbb{E}[X_t|U_{0}^t,A,B])].
\end{equation}
We will first show that $f(t)$ is right continuous.
Let $\tilde{\delta}_n$ be a non-increasing positive subsequence approaching $0$ as $n\to\infty$. Define the following (suppressing the time index $t$)
\begin{align}
\mathcal{H}_n&=\F_{(t+\tilde{\delta}_n)}^U\vee\sigma(A)\vee\sigma(B) \\
\mathsf{X}_n&=X_{t+\tilde{\delta}_n}.
\end{align}
Since the sample paths of $X_t$ are right-continuous 
$$\lim_{n\to\infty}\mathsf{X}_n\to X_t$$ and $\mathcal{H}_{1}\supset\mathcal{H}_{2}\supset\dots$, we have the following equalities $P$-a.s. 
\begin{align}
\lim_{n\to\infty}\mathbb{E}[X_{t+\tilde{\delta}_n}|U_0^{t+\tilde{\delta}_n},A,B]&\stackrel{(a)}{=}\lim_{n\to\infty}\mathbb{E}[\mathsf{X}_n|\mathcal{H}_n]\nn
&\stackrel{(b)}{=}\mathbb{E}\left[X_t \bigg|\bigcap_n\mathcal{H}_n\right]\nn
&=\mathbb{E}\left[X_t\middle|\bigcap_{\epsilon> 0} \F_{t+\epsilon}^U\vee\sigma(A)\vee\sigma(B)\right] \nn
&\stackrel{(c)}{=}\mathbb{E}[X_t|\F_{t}^U\vee\sigma(A)\vee\sigma(B)] \nn
&=\mathbb{E}[X_t|U_0^t,A,B].
\end{align}
Here, (a) is due to the definition of $\mathsf{X}_n$ and $\mathcal{H}_n$, \\*
(b) is due to the  backwards analogue of the  dominated convergence theorem for conditional expectation \cite[Exercise 5.6.2, p. 265]{Durrett} (recall that $X_t$  is bounded), and \\*
(c) is due to the right continuity of the filtration $\F_{t}^U\vee\sigma(A)\vee\sigma(B)$ \cite[Theorem T25, Appendix A2, p. 304]{Bremaud}. \\*
Since $\phi_u(x)$ is a continuous function and $X_t$ is a bounded random variable 
$$
\lim_{\tilde{\delta}_n\to 0^+}\mathbb{E}[\phi_u(\mathbb{E}[X_{t+\tilde{\delta}_n}|U_0^{t+\tilde{\delta}_n},A,B])]=\mathbb{E}[\phi_u(\mathbb{E}[X_t|U_0^t,A,B])],
$$
and hence 
$$
\lim_{\delta\to 0^+}\mathbb{E}[\phi_u(\mathbb{E}[X_{t+\delta}|U_0^{t+\delta},A,B])]=\mathbb{E}[\phi_u(\mathbb{E}[X_t|U_0^t,A,B])].
$$
Similarly,
$$
\lim_{\delta\to 0^+}\mathbb{E}[\phi_u(\mathbb{E}[X_{t+\delta}|U_0^{t+\delta},B])]=\mathbb{E}[\phi_u(\mathbb{E}[X_t|U_0^t,B])].
$$
Since $(A,B)\rightleftarrows X_0^T \rightleftarrows (U_0^{t},U_t^{t+\delta})$ and $U_0^t \rightleftarrows X_0^T \rightleftarrows U_t^{t+\delta}$ are Markov chains, \cite[Proposition 6.8, p. 111]{Kallenberg} implies $(A,B,U_0^t)\rightleftarrows X_0^T \rightleftarrows U_t^{t+\delta}$ is also a Markov chain.  Taking $t_1=t$, $t_2=t+\delta$, Theorem \ref{Th:Mu_inf} yields
\begin{align}
\lim_{\delta\to 0^+}\frac{1}{\delta}I\left(A;U_t^{t+\delta}\middle| U_0^t,B\right)&=\lim_{\delta\to 0^+}\frac{1}{\delta}\int_t^{t+\delta}\mathbb{E}[\phi_u(\mathbb{E}[X_s|U_t^{s},U_0^t,A,B])]-\mathbb{E}[\phi_u(\mathbb{E}[X_s|U_t^{s},U_0^t,B])]  \,ds \nn
&=\mathbb{E}[\phi_u(\mathbb{E}[X_t|U_0^t,A,B])]-\mathbb{E}[\phi_u(\mathbb{E}[X_t|U_0^t,B])],
\end{align}
 where the last equality is due to the fact that if $f(x)$ is right continuous at $t$, then 
$$
\lim_{\delta\to 0^+}\frac{1}{\delta}\int_t^{t+\delta} f(s)\,ds=f(t).
$$
Let $\tilde{U}_0^T$ to be the time-reversed  $U_0^T$ process. Then $\tilde{U}_0^T$ is a doubly-stochastic Poisson process  with rate process $\{\tilde{X}_t=X_{(T-t)-},\tin\}$, and
\begin{align}
\lim_{\delta\to 0^+}\frac{1}{\delta}I\left(A;U_{t-\delta}^{t}\middle| U_t^T,B\right)&=\lim_{\delta\to 0^+}\frac{1}{\delta}I\left(A;\tilde{U}_{T-t}^{T-t+\delta}\middle| \tilde{U}^{T-t}_0,B\right)\nn
&=\mathbb{E}[\phi_u(\mathbb{E}[\tilde{X}_{(T-t)-}|\tilde{U}_0^{T-t},A,B])]-\mathbb{E}[\phi_u(\mathbb{E}[\tilde{X}_{T-t}|\tilde{U}_0^{T-t},B])]\nn
&=\mathbb{E}[\phi_u(\mathbb{E}[X_{t-}|U_t^T,A,B])]-\mathbb{E}[\phi_u(\mathbb{E}[X_{t-}|U_t^T,B])].
\end{align}
\end{IEEEproof}

\begin{IEEEproof}[Proof of Lemma \ref{Le:Bounded}]
We have
\begin{align}
\frac{1}{\delta}I\left(A;U_s^{s+\delta}\middle| U_0^s,B\right)&=\frac{1}{\delta}\int_s^{s+\delta}\mathbb{E}[\phi_u(\mathbb{E}[X_r|U_0^{r},A,B])]-\mathbb{E}[\phi_u(\mathbb{E}[X_r|U_0^{r},B])]  \,dr \nn
&\le 2\phi_u^*,
\end{align}
where $\phi_u^*=\max\limits_{0\le x \le 1}|\phi_u(x)|$. 
The second part of the lemma follows similarly.
\end{IEEEproof}

\begin{IEEEproof}[Proof of Lemma \ref{Le:Ch_Rule_Cont}]
Consider
\begin{align}
I(A;U_{0}^{t}|B)&\stackrel{(a)}{=}\int_{0}^{t}\mathbb{E}[\phi_u(\mathbb{E}[X_s|U_0^s,A,B])]-\mathbb{E}[\phi_u(\mathbb{E}[X_s|U_0^s,B])]  \,ds \nn
&\stackrel{(b)}{=}\int_{0}^{t}\lim_{\delta\to 0^+}\frac{1}{\delta}I\left(A;U_s^{s+\delta}\middle| U_0^s,B\right)\,ds\nn
&\stackrel{(c)}{=}\lim_{\delta\to 0^+}\frac{1}{\delta}\int_{0}^{t}I\left(A;U_s^{s+\delta}\middle| U_0^s,B\right)\,ds.
\end{align}
Here,
(a) is due to Theorem \ref{Th:Mu_inf},\\*
(b) is due to Lemma \ref{Le:Ch_Rule}, and \\*
(c) is due to  Lemma \ref{Le:Bounded} and the dominated convergence theorem.

The proof of the second part of the lemma follows similarly. 
\end{IEEEproof}

\begin{IEEEproof}[Proof of Lemma \ref{Le:Conv}]
The existence of $\mathfrak{i}(X_0^{T_n};Y_0^{T_n})$ and $\mathfrak{i}(V_0^{T_n};Z_0^{T_n})$ is due to Lemma \ref{Le:Abs_cont}. The  existence of $\mathfrak{i}(X_0^{T_n};Y_0^{T_n}|V_0^{T_n})$ is discussed in a later part of this proof.
We will use the measure $\tilde{P}$ as defined in Theorem \ref{Th:Mu_inf}. Using Theorem \ref{Th:Mu_inf} we have $P^{V_0^{T_n},X_0^{T_n},Z_{0}^{T_n}}$-a.s.
\begin{align}
\frac{1}{T_n}\log\left(\frac{dP^{V_0^{T_n},Z_{0}^{T_n}}}{d\tilde{P}^{V_0^{T_n},Z_{0}^{T_n}}}\right) &=\frac{1}{T_n}\int_{0}^{T_n}\log(a_z\Pi_t+\lambda_z)d{Z}_t+1-(a_z\Pi_t+\lambda_z)\,dt&\label{EQ:RN1}\nn
 &=\frac{1}{{n\tau}}\sum_{i=1}^n\int_{(i-1)\tau}^{i\tau}\log(a_z\Pi_t+\lambda_z)\,d{Z}_t^{(i)}+1-(a_z\Pi_t+\lambda_z)\,dt,
\end{align} 
where $\left\{{Z}_t^{(i)}, t\in\left[(i-1)\tau, i\tau\right]\right\}$ is the point process corresponding to $Z_{(i-1)\tau}^{i\tau}$, and for $\tin$,
\begin{align*}
\Pi_t=E[X_t|Z_{0}^t,{V}_0^{T_n}], \quad P^{V_0^{T_n},X_0^{T_n},Z_{0}^{T_n}}\text{-a.s.}
\end{align*}
Let 
\begin{align}
\Psi_i^{(1)}=\frac{1}{\tau}\int_{(i-1)\tau}^{i\tau} \log(a_z\Pi_t+\lambda_z)\,d{Z}^{(i)}_t,
\end{align}
then $\Psi_i^{(1)}$, for $i=1,2,\dots, n$ are independent and identically distributed with
\begin{align}
\E[|\Psi_1^{(1)}|]=&\frac{1}{\tau}\E\left[\left|\int_{0}^{\tau} \log(a_z\Pi_t+\lambda_z)\,d{Z}_t\right|\right]\nn
&\le \frac{1}{\tau}\E\left[\int_{0}^{\tau} \left|\log(a_z\Pi_t+\lambda_z)\right|\,d{Z}_t\right]\nn
&=\frac{1}{\tau}\int_{0}^{\tau}\E\left[ \left|\phi_z\left(\Pi_t\right)\right|\right]\,dt\nn
&\le \phi_z^* < \infty,
\end{align}
where $\phi_z^*=\max\limits_{0\le x \le 1}\phi_z(x)$, and we have used the fact that the $(P,\sigma(\bar{V}_1)\vee\F_t^Z:\tin)$-intensity of $Z_0^{T_n}$ is $a_z\Pi_t+\lambda_z$ (Lemma \ref{Le:MC_Int}). Thus by the strong law of large numbers \cite[Theorem 4.23, p.73]{Kallenberg}
\begin{align}
\frac{1}{n}\sum_{i=1}^{n}\Psi_i^{(1)}\to\E[\Psi_1^{(1)}]=\frac{1}{\tau}\int_{0}^{\tau}\E\left[\phi_z\left(\E[X_t|Z_{0}^t,\bar{V}_1]\right)\right]\,dt
\end{align}
almost surely. Now let
\begin{align*}
\Psi_i^{(2)}=\frac{1}{\tau}\int_{(i-1)\tau}^{i\tau} a_z\Pi_t+\lambda_z\,dt,
\end{align*}
for which the law of large numbers gives
\begin{align}
\frac{1}{n}\sum_{i=1}^{n}\Psi_i^{(2)}\xrightarrow{\as}\E[\Psi_1^{(2)}]=\frac{1}{\tau}\int_{0}^{\tau}a_z\E[X_t]+\lambda_z\,dt.
\end{align}
Thus
\begin{align}
\frac{1}{T_n}\log\left(\frac{dP^{V_0^{T_n},Z_{0}^{T_n}}}{d\tilde{P}^{V_0^{T_n},Z_{0}^{T_n}}}\right)\xrightarrow{\as}\frac{1}{\tau}\int_{0}^{\tau}\E\left[\phi_z\left(\E[X_t|Z_{0}^t,\bar{V}_1]\right)\right]+1-(a_z\E[X_t]+\lambda_z)\,dt.
\end{align} 
Similarly $P^{V_0^{T_n},X_0^{T_n},Z_{0}^{T_n}}$-a.s.
\begin{align}
\frac{1}{T_n}\log\left(\frac{dP^{Z_{0}^{T_n}}}{dP_0^{Z_{0}^{T_n}}}\right)
&\xrightarrow{\as}\frac{1}{\tau}\int_{0}^{\tau}\E\left[\phi_z\left(\E[X_t|Z_{0}^t]\right)\right]+1-(a_z\E[X_t]+\lambda_z)\,dt.
\end{align} 
This gives $P^{V_0^{T_n},X_0^{T_n},Z_{0}^{T_n}}$-a.s.
\begin{align}
\frac{1}{T_n}\ii(V_0^{T_n};Z_0^{T_n})&=\log\frac{dP^{V_0^{T_n},Z_0^{T_n}}}{d(P^{V_0^{T_n}}\times P^{Z_0^{T_n}})}\nn
&=\frac{1}{T_n}\log\left(\frac{dP^{V_0^{T_n},Z_{0}^{T_n}}}{d\tilde{P}^{V_0^{T_n},Z_{0}^{T_n}}}\right)-\frac{1}{T_n}\log\left(\frac{dP^{Z_{0}^{T_n}}}{d{P_0}^{Z_{0}^{T_n}}}\right)\nn
&\xrightarrow{\as}\frac{1}{\tau}\int_{0}^{\tau}\E\left[\phi_z\left(\E[X_t|Z_{0}^t,\bar{V}_1]\right)\right]-\E\left[\phi_z\left(\E[X_t|Z_{0}^t]\right)\right]\,dt \nn
&=\frac{1}{\tau}I(\bar{V}_1;Z_0^{\tau})
\end{align}
as $n\to\infty$, and we have used Theorem \ref{Th:Mu_inf}.  From Lemma \ref{Le:Ch_Rule}
\begin{align}
\lim_{\tau \to 0^{+}}\frac{1}{\tau}I(\bar{V}_1;Z_0^{\tau})
&=\E\left[\phi_z\left(\E[X_0|\bar{V}_1]\right)\right]-\phi_z\left(\E[X_0]\right).
\end{align}
Thus given any $\epsilon>0$, we can choose  $\bar{\tau}$ such that 
\begin{align}
\left|\frac{1}{\tau^*}I(\bar{V}_1;Z_0^{\tau^*})-\left(\E\left[\phi_z\left(\E[X_0|\bar{V}_1]\right)\right]-\phi_z\left(\E[X_0]\right)\right) \right|\le \frac{\epsilon}{2},
\end{align}
and then choosing $N$ large enough we can ensure that 
\begin{align}
P\left(\left|\frac{1}{T_N}\ii(V_0^{T_N};Z_0^{T_N})-\left(\E\left[\phi_z\left(\E[X_0|\bar{V}_1]\right)\right]-\phi_z\left(\E[X_0]\right)\right)\right| > \epsilon \right) \le \epsilon.
\end{align}

Note that $\mathcal{V}_0^{T_n}$ and  $\mathcal{X}_0^{T_n}$ here are effectively finite alphabets. For the space $(\mathcal{N}_0^{T_n},\Fk^Y)$, the $\sigma$-field $\Fk^Y$ is the restriction of the $\sigma$-field generated by the Skorohod topology on $D[0, 1]$ to $\mathcal{N}_0^{T_n}$. This makes $(\mathcal{N}_0^{T_n},\Fk^Y)$ a standard space \cite[Theorem 12.2, p. 128]{Billingsley} and  \cite[Section 1.5, p. 12]{Gray}.   Consider
\begin{align}
I(V_0^{T_N},X_0^{T_N};Y_0^{T_n})&=I(X_0^{T_n};Y_0^{T_N})+I(V_0^{T_n};Y_0^{T_N}|X_0^{T_N})\nn
&=I(X_0^{T_n};Y_0^{T_N})<\infty.
\end{align}
This gives ${P}^{V_0^{T_n},X_0^{T_n},Y_{0}^{T_n}}\ll {P}^{V_0^{T_n},X_0^{T_n}}\times P^{Y_{0}^{T_n}}$.  Thus from \cite[Corollary 5.5.3, p. 125]{Gray},  $\ii(X_0^{T_n};Y_0^{T_n}|V_0^{T_n})$ exists and ${P}^{V_0^{T_n},X_0^{T_n},Y_{0}^{T_n}}$-a.s. satisfies
\begin{align}
\frac{1}{T_n}\ii(X_0^{T_n};Y_0^{T_n}|V_0^{T_n})&=\frac{1}{T_n}\ii(V_0^{T_n},X_0^{T_n};Y_0^{T_n})-\frac{1}{T_n}\ii(V_0^{T_n};Y_0^{T_n}).
\end{align}
Here, we have used the fact that  since $\frac{1}{T_n}\E[|\ii(V_0^{T_n};Y_0^{T_n})|]<\infty $, $\frac{1}{T_n}\ii(V_0^{T_n};Y_0^{T_n})$ is ${P}^{V_0^{T_n},X_0^{T_n},Y_{0}^{T_n}}$-a.s. finite. Also ${P}^{X_0^{T_n},Y_{0}^{T_n}}\ll {P}^{X_0^{T_n}}\times P^{Y_{0}^{T_n}}$ (since $I(X_0^{T_n};Y_{0}^{T_n})<\infty$), and $V_0^{T_n} \rightleftarrows X_0^{T_n} \rightleftarrows Y_0^{T_n}$ being a Markov chain, \cite[Corollary 5.5.4, p.126]{Gray} yields 
\begin{align*}
\ii(V_0^{T_n},X_0^{T_n};Y_0^{T_n})=\ii(X_0^{T_n};Y_0^{T_n}), \quad P^{V_0^{T_n},X_0^{T_n},Y_{0}^{T_n}}\text{-a.s.} 
\end{align*}
Since $P^{X_0^{T_n},Y_0^{T_n}}$-a.s.
\begin{align*}
\ii(X_0^{T_n};Y_0^{T_n})&=\log\left(\frac{dP^{X_0^{T_n},Y_0^{T_n}}}{d\left(P^{X_0^{T_n}}\times P^{Y_0^{T_n}}\right)}\right)\nn
&=\log\left(\frac{dP^{X_0^{T_n},Y_0^{T_n}}}{d\tilde{P}^{X_0^{T_n},Y_0^{T_n}}}\right)-\log\left(\frac{dP^{Y_0^{T_n}}}{dP_0^{Y_0^{T_n}}}\right),
\end{align*}
we have from Theorem \ref{Th:Mu_inf}, $P^{V_0^{T_n},X_0^{T_n},Y_0^{T_n}}$-a.s.
\begin{align*}
\frac{1}{T_n}\log\left(\frac{dP^{{X_0^{T_n}},Y_0^{T_n}}}{d\tilde{P}^{{X_0^{T_n}},Y_0^{T_n}}}\right)&=\frac{1}{T_n}\int_{0}^{T_n}\log(a_yX_{t}+\lambda_y)d\,{Y}_t+1-(a_yX_{t}+\lambda_y)\,dt \nn
&\xrightarrow{a.s.}\frac{1}{\tau}\int_{0}^{\tau}\E\left[\phi_y\left(X_t\right)\right]+1-(a_y\E[X_t]+\lambda_y)\,dt,
\end{align*}
where the a.s. convergence can be shown by using an argument similar to that used for the second user. Similarly for the second term, $P^{V_0^{T_n},X_0^{T_n},Y_0^{T_n}}$-a.s.,
\begin{align*}
\frac{1}{T_n}\log\left(\frac{dP^{Y_0^{T_n}}}{dP_0^{Y_0^{T_n}}}\right)&=\frac{1}{T_n}\int_{0}^{T_n}\log(a_y\Pi^\prime_t+\lambda_y)d{Y}_t+1-(a_y\Pi^\prime_t+\lambda_y)\,dt \nn
&\xrightarrow{a.s.}\frac{1}{\tau}\int_{0}^{\tau}\E\left[\phi_y\left(\E[X_t|Y_0^t]\right)\right]+1-(a_y\E[X_t]+\lambda_y)\,dt,
\end{align*}
where $\Pi^\prime_t=\E[X_t|Y_0^t]$ $P^{V_0^{T_n},X_0^{T_n},Y_0^{T_n}}$-a.s.
Hence we have
\begin{align*}
\frac{1}{T_n}\ii(X_0^{T_n};Y_0^{T_n})
&\xrightarrow{\as}\frac{1}{\tau}\int_{0}^{\tau}\E\left[\phi_y(X_t)\right]-\E\left[\phi_y\left(\E[X_t|Y_{0}^t]\right)\right]\,dt\nn
&=\frac{1}{\tau}I(X_0^\tau;Y_0^\tau)=\frac{1}{\tau}I(X_0;Y_0^\tau),
\end{align*}
where we have used the fact that $X_0^\tau$ is constant over the interval $[0,\tau)$ and Theorem \ref{Th:Mu_inf}.
From Lemma \ref{Le:Ch_Rule}
\begin{align*}
\lim_{\tau\to0^+}\frac{1}{\tau}I(X_0;Y_0^\tau)=\E\left[\phi_y(X_0)\right]-\phi_y\left(\E[X_0]\right).
\end{align*}

Also, similar to the second receiver, we can show that for a given $\epsilon>0$ there exists $N$ and $\bar{\tau}$ such that $n\ge N$ and $\tau \le \bar{\tau}$ implies that
\begin{align}
P\left(\left|\frac{1}{T_n}\mathfrak{i}(V_0^{T_n};Y_0^{T_n})-\left(\E\left[\phi_y\left(\E[X_0|\bar{V}_1]\right)\right]-\phi_y(\E[X_0])\right)\right|>\epsilon\right)\le\epsilon
\end{align}
Since $P^{V_0^{T_n},X_0^{T_n},Y_{0}^{T_n}}$-a.s.
\begin{align*}
\frac{1}{T_n}\ii(X_0^{T_n};Y_0^{T_n}|V_0^{T_n})&=\frac{1}{T_n}\ii(X_0^{T_n};Y_0^{T_n})-\frac{1}{T_n}\ii(V_0^{T_n};Y_0^{T_n}).
\end{align*}
Thus for given $\epsilon>0$ there exists $N$ and $\bar{\tau}$ such that $n\ge N$ and $\tau \le \bar{\tau}$ implies that
\begin{align}
P\left(\left|\frac{1}{{T_n}}\mathfrak{i}(X_0^{T_n};Y_0^{T_n}|V_0^{T_n})-\left(\E\left[\phi_y(X_0)\right]-\E\left[\phi_y\left(\E[X_0|\bar{V}_1]\right)\right]\right)\right|>\epsilon\right)\le\epsilon.
\end{align}
\end{IEEEproof}

\begin{IEEEproof}[Proof of Lemma \ref{Le:ln_inq}]
Note that
\begin{eqnarray*}
&&\int_0^T\mathbb{E}[\phi_z(\mathbb{E}[X_t|M,Y_0^t])] - \mathbb{E}[\phi_z(\mathbb{E}[X_t|M,Z_t^T])] \,dt\\
&\stackrel{(a)}{=}&\int_0^T\mathbb{E}[\phi_y(\mathbb{E}[X_t|M,Y_0^t])] - \mathbb{E}[\phi_z(\mathbb{E}[X_t|M,Z_t^T])]\,dt \nonumber\\
&&-\int_0^T\mathbb{E}[\phi_y(\mathbb{E}[X_t|M,Y_0^t])]-\mathbb{E}[\phi_z(\mathbb{E}[X_t|M,Y_0^t])]\,dt\\
&\stackrel{(b)}{=}&\int_0^T\mathbb{E}[\phi_y(\mathbb{E}[X_t|M,Y_0^t,Z_t^T])] - \mathbb{E}[\phi_z(\mathbb{E}[X_t|M,Y_0^t,Z_t^T])]\,dt \nonumber\\
&&-\int_0^T\mathbb{E}[\phi_y(\mathbb{E}[X_t|M,Y_0^t])]-\mathbb{E}[\phi_z(\mathbb{E}[X_t|M,Y_0^t])]\,dt\\
&\stackrel{(c)}{=}&\int_0^T\mathbb{E}[\Phi(\mathbb{E}[X_t|M,Y_0^t,Z_t^T])]-\mathbb{E}[\Phi(\mathbb{E}[X_t|M,Y_0^t])]\,dt\\
&\stackrel{(d)}{\geq}&0. \nonumber
\end{eqnarray*}
In (a) we have added and subtracted a term,\\*
(b) is due to Theorem \ref{Thm:Integral_Identity},\\* 
 (c) is due to the definition of $\Phi(x)$, and \\*
 (d) is due to convexity of $\Phi(x)$ and Jensen's inequality. 
\end{IEEEproof}

\begin{IEEEproof}[Proof of Lemma \ref{Le:Conv1}]
In this case we have
\begin{align*}
\E\left[\phi_y(X_0)\right]-\phi_y\left(\E[X_0]\right)
&=(\alpha p+(1-\alpha)q)\phi_y(1)+(\alpha (1-p)+(1-\alpha)(1-q))\phi_y(0)-\phi_y(\alpha p+(1-\alpha)q)\nn
&=C_y+\tilde{C}_y.
\end{align*}
And
\begin{align*}
\E\left[\phi_z\left(\E[X_0|\bar{V}_1]\right)\right]-\phi_z\left(\E[X_0]\right)
&=\alpha\phi_z(p)+(1-\alpha)\phi_z(q)-\phi_z(\alpha p+(1-\alpha)q)\nn
&=C_z.
\end{align*}
Similarly,
\begin{align*}
\E\left[\phi_y(X_0)\right]-\E\left[\phi_y\left(\E[X_0|\bar{V}_1]\right)\right]\
&=(\alpha p+(1-\alpha)q)\phi_y(1)+(\alpha (1-p)+(1-\alpha)(1-q))\phi_y(0)\nn
&\quad-\alpha\phi_y(p)-(1-\alpha)\phi_y(q)\nn
&=C_y.
\end{align*}
Now applying Lemma \ref{Le:Conv} proves the statement of the lemma. 
\end{IEEEproof}

\bibliographystyle{IEEEtran}
\bibliography{Poisson}

\begin{thebibliography}{10}
\providecommand{\url}[1]{#1}
\csname url@samestyle\endcsname
\providecommand{\newblock}{\relax}
\providecommand{\bibinfo}[2]{#2}
\providecommand{\BIBentrySTDinterwordspacing}{\spaceskip=0pt\relax}
\providecommand{\BIBentryALTinterwordstretchfactor}{4}
\providecommand{\BIBentryALTinterwordspacing}{\spaceskip=\fontdimen2\font plus
\BIBentryALTinterwordstretchfactor\fontdimen3\font minus
  \fontdimen4\font\relax}
\providecommand{\BIBforeignlanguage}[2]{{%
\expandafter\ifx\csname l@#1\endcsname\relax
\typeout{** WARNING: IEEEtran.bst: No hyphenation pattern has been}%
\typeout{** loaded for the language `#1'. Using the pattern for}%
\typeout{** the default language instead.}%
\else
\language=\csname l@#1\endcsname
\fi
#2}}
\providecommand{\BIBdecl}{\relax}
\BIBdecl

\bibitem{Personick}
S.~Personick, ``Receiver design for digital fiber optic communication systems,
  {II},'' \emph{The Bell System Technical Journal}, vol.~52, no.~6, pp.
  875--886, July 1973.

\bibitem{Mazo}
J.~Mazo and J.~Salz, ``On optical data communication via direct detection of
  light pulses,'' \emph{The Bell System Technical Journal}, vol.~55, no.~3, pp.
  347--369, Mar 1976.

\bibitem{Kabanov}
Y.~Kabanov, ``The capacity of a channel of the {P}oisson type,'' \emph{Theory
  of Probabilty and Applications}, vol.~23, pp. 143--147, 1978.

\bibitem{Davis}
M.~Davis, ``Capacity and cutoff rate for {P}oisson-type channels,'' \emph{IEEE
  Transactions on Information Theory}, vol.~26, no.~6, pp. 710--715, Nov 1980.

\bibitem{Wyner}
A.~Wyner, ``Capacity and error exponent for the direct detection photon
  channel{-} {P}art {I} and {II},'' \emph{IEEE Transactions on Information
  Theory}, vol.~34, no.~6, pp. 1449--1461, Nov 1988.

\bibitem{Shannon}
C.~Shannon, ``A mathematical theory of communication,'' \emph{The Bell System
  Technical Journal}, vol.~27, no.~3, pp. 379--423, July 1948.

\bibitem{Bergmans}
P.~Bergmans, ``Random coding theorem for broadcast channels with degraded
  components,'' \emph{IEEE Transactions on Information Theory}, vol.~19, no.~2,
  pp. 197--207, Mar 1973.

\bibitem{Korner}
J.~K{\"o}rner and K.~Marton, ``Comparison of two noisy channels,'' \emph{Topics
  in information theory}, no.~16, 1977.

\bibitem{Gamal1}
A.~El~Gamal, ``The capacity of a class of broadcast channels,'' \emph{IEEE
  Transactions on Information Theory}, vol.~25, no.~2, pp. 166--169, Mar 1979.

\bibitem{Kim}
H.~Kim, B.~Nachman, and A.~El~Gamal, ``Superposition coding is almost always
  optimal for the {P}oisson broadcast channel,'' \emph{IEEE Transactions on
  Information Theory}, vol.~62, no.~4, pp. 1782--1794, April 2016.

\bibitem{WYNER197851}
A.~Wyner, ``A definition of conditional mutual information for arbitrary
  ensembles,'' \emph{Information and Control}, vol.~38, no.~1, pp. 51 -- 59,
  1978.

\bibitem{Gelfand}
I.~M. Gel'fand and A.~M. Yaglom, ``Computation of the amount of information
  about a~stochastic function contained in another such function,''
  \emph{Uspekhi Mat. Nauk}, vol.~12, no.~1, pp. 3--52, 1957.

\bibitem{Gray}
R.~M. {G}ray, \emph{Entropy and Information Theory}.\hskip 1em plus 0.5em minus
  0.4em\relax Springer-Verlag, 1990.

\bibitem{Billingsley}
P.~Billingsley, \emph{Convergence of Probability Measures}, 2nd~ed.\hskip 1em
  plus 0.5em minus 0.4em\relax Wiley Series in Probability and Statistics,
  1999.

\bibitem{Bremaud}
P.~{B}r\'{e}maud, \emph{Point Procceses and Queues: Martingale Dynamics}.\hskip
  1em plus 0.5em minus 0.4em\relax Springer-Verlag, 1981.

\bibitem{Kallenberg}
O.~Kallenberg, \emph{Foundations of Modern Probability}, 2nd~ed.\hskip 1em plus
  0.5em minus 0.4em\relax Springer-Verlag, New York, 2002.

\bibitem{Liptser}
R.~S. Liptser and A.~N. Shiryaev, \emph{Statistics of Random Processes II},
  2nd~ed.\hskip 1em plus 0.5em minus 0.4em\relax Springer-Verlag Berlin
  Heidelberg, 2001.

\bibitem{Csiszar}
I.~Csisz\'{a}r and J.~K\"{o}rner, ``Broadcast channels with confidential
  messages,'' \emph{IEEE Transactions on Information Theory}, vol.~24, no.~3,
  pp. 339--348, May 1978.

\bibitem{Lapidoth}
A.~Lapidoth, I.~Telatar, and R.~Urbanke, ``On wide-band broadcast channels,''
  \emph{IEEE Transactions on Information Theory}, vol.~49, no.~12, pp.
  3250--3258, Dec 2003.

\bibitem{Nair}
C.~Nair, ``Capacity regions of two new classes of two-receiver broadcast
  channels,'' \emph{IEEE Transactions on Information Theory}, vol.~56, no.~9,
  pp. 4207--4214, Sept 2010.

\bibitem{Salehi}
I.~Csisz{\'a}r and J.~K{\"o}rner, \emph{Information Theory: Coding Theorems for
  Discrete Memoryless Systems}, 2nd~ed.\hskip 1em plus 0.5em minus 0.4em\relax
  Cambridge University Press, 2011.

\bibitem{Laourine}
A.~Laourine and A.~B. Wagner, ``The degraded {P}oisson wiretap channel,''
  \emph{IEEE Transactions on Information Theory}, vol.~58, no.~12, pp.
  7073--7085, Dec 2012.

\bibitem{Dellacherie}
C.~{D}ellacherie and P.~A. Meyer, \emph{Probabilities and Potential B: Theory
  of Martingales}, ser. North-Holland Mathematics Studies.\hskip 1em plus 0.5em
  minus 0.4em\relax North-Holland, 1982, vol.~72.

\bibitem{Durrett}
R.~Durrett, \emph{Probabilty Theory and Examples}, 4th~ed.\hskip 1em plus 0.5em
  minus 0.4em\relax Cambridge University Press, 2010.

\end{thebibliography}
\end{document}